\documentclass[graybox, envcountchap]{svmult}

\usepackage{mathptmx}         
\usepackage{amsmath}
\usepackage{amssymb}
\usepackage{ulem}
\usepackage{color}
\usepackage{helvet}           
\usepackage{courier}          
\usepackage{dirtree}

\usepackage{makeidx}          
\usepackage{graphicx}         
\usepackage{subfig}

\usepackage{multicol}         
\usepackage[bottom]{footmisc} 
\usepackage{adjustbox}
\usepackage{multirow}

\usepackage{hyperref}         
\hypersetup{colorlinks=true,urlcolor=blue}

\usepackage[misc]{ifsym}

\makeindex             

\allowdisplaybreaks

\def\p{\partial}
\def\Lie{{\cal L}}

\begin{document}

\title*{Solving the Einstein Equations Numerically}
\author{David Hilditch}
\institute{David Hilditch (\Letter) \at CENTRA, Departamento de
  Física, Instituto Superior Técnico – IST, Universidade de Lisboa –
  UL, Avenida Rovisco Pais 1, 1049-001 Lisboa, Portugal,
  \email{david.hilditch@tecnico.ulisboa.pt}}

\maketitle

\abstract{There are many complementary approaches to the construction
  of solutions to the field equations of general relativity. Among
  these, numerical approximation offers the only possibility to
  compute a variety of dynamical spacetimes, and so has come to play
  an important role for theory and experiment alike. Presently we give
  a brief introduction to this, the science of numerical
  relativity. We discuss the freedom in formulating general relativity
  as an initial (boundary) value problem. We touch on the fundamental
  concepts of well-posedness and gauge freedom and review the standard
  computational methods employed in the field. We discuss the physical
  interpretation of numerical spacetime data and end with an overview
  of a number of 3d codes that are either in use or under active
  development.}

\section{Overview}
\label{Sec:Overview}

This book is primarily concerned with advanced computational methods
for general relativistic magnetohydrodynamics (GRMHD) and the
solutions they produce. It is therefore assumed at the outset that the
reader already has a solid grasp of the fundamentals of general
relativity (GR)~\cite{Wal84}. There are, however, many practitioners
working on GRMHD who enter the subject through computational
astrophysics, where it is common to solve the field equations of GRMHD
with a given, fixed, spacetime metric. Following on from the previous
Chapter~\cite{MizRez24}, in which a concise overview to GRMHD was
given, our purpose here is to give a brief introduction to the field
of numerical relativity (NR), where the metric becomes an additional
dependent variable. There are excellent
textbooks~\cite{Alc08,BauSha10,Gou12,Shi16,BauSha21}, which are highly
recommended for those seeking instead a complete introduction to the
topic.

Starting with the basics, we recall that in GR, gravitation is
described geometrically within the spacetime manifold~$M$ as the
curvature of the Levi-Civita covariant derivative~$\nabla_a$
associated with the Lorentzian metric~$g_{ab}$, which we take to have
mostly plus signature. The goal of the relativist is then to
understand the solution space of Einstein's equations, the field
equations of the theory,
\begin{align}
  {}^{{\tiny (4)}}G_{ab}\equiv {}^{{\tiny (4)}}R_{ab}
  -\tfrac{1}{2}g_{ab} {}^{{\tiny (4)}}R
  = 8 \pi T_{ab}\,,\label{eqn:EEs}
\end{align}
expressed here and throughout in units with~$G=c=1$. Latin indices
starting from the beginning of the alphabet will be abstract, and we
shall work in the physical~$3+1$ dimensions. The label~$(4)$ denotes
then the spacetime curvature, distinct from the spatial curvature
introduced below. The generalization of the results presented in this
Chapter to higher or lower spacetime dimensions is
straightforward. The Ricci tensor and scalar are given, respectively
by,
\begin{align}
  {}^{{\tiny (4)}}R_{ab}={}^{{\tiny (4)}}R_{acb}{}^c\,,
  \qquad {}^{{\tiny (4)}}R=g^{ab}\,{}^{{\tiny (4)}}R_{ab}\,,
\end{align}
with the inverse metric denoted as usual by~$g^{ab}$. Introducing
coordinates~$x^\mu$ and writing out the field equations in the
associated coordinate basis we obtain a system of coupled nonlinear
partial differential equations (PDEs) whose solution space we can
first try to understand with various simplifying strategies. Under
strong symmetry assumptions, for instance, we may be able to find
exact analytic solutions. The most important of these for the
astrophysics of black holes is the Kerr solution. If we are interested
in physical scenarios that are close to a known solution, we may
linearize about that background and solve these simplified equations,
whose solutions nevertheless accurately describe the physics of
interest. This approach is fundamental to understand the propagation
of weak gravitational waves far from their original source. More
generally, when we encounter a problem with two or more very different
scales we can use the smaller of the two as an expansion
parameter. This strategy is used in the self-force treatment of
extreme mass-ratio compact binary systems and in the Post-Newtonian
expansion of the field equations, which has proven highly successful
in the computation of the inspiral motion of compact binary systems.
Turning instead to produce and understand solutions to the full field
equations without simplifying assumptions, methods of pure
mathematical analysis are absolutely fundamental to obtain a rigorous
picture. Yet the general message we must absorb from the PDE
literature is that such pure analysis will not provide solutions {\it
  in hand}. In many physical applications however, we do need the
solution itself. To overcome this we turn to NR: the use of numerical
methods to find approximate solutions to the field equations that
converge to solutions of the continuum equations in the limit of
infinite resolution. We will see that the requirement of convergence
guides much of the theory in the topic. The most important application
of NR to astrophysics is in the computation of the late stages of
compact binary inspiral and merger.

In the following we give an overview of the basics of NR. We begin in
Section~\ref{Sec:Formulations} by reviewing the formulation of GR as
an initial value and initial boundary value problem. We discuss
well-posedness and the most popular free-evolution formulations of
GR. In Section~\ref{Sec:Interpretation} the post-processing used to
give physical interpretation to numerically constructed spacetime data
is discussed. Finally, in Section~\ref{Sec:Numerical_Methods} we give
an introduction to the standard methods used in NR, and finally give
an overview of the 3d codes in use.

\section{Formulations of GR}
\label{Sec:Formulations}

Perhaps the single most important insight of special and general
relativity is that space and time should be viewed as a unified
entity. Using this insight to express field theories of interest
results in a minimal and elegant formulation of their field equations.
Yet such a spacetime formulation is not optimally adapted to the
treatment of the initial and initial boundary value problems that
necessarily underpin any understanding of dynamics within the
theory. Textbook numerical methods for time evolution moreover rely on
a clear separation of temporal and spatial derivatives. In this
section, focusing primarily on the evolution of the metric in GR, we
review the~$3+1$ decomposition, which allows this
separation. Afterwards we move on to discuss free-evolution,
well-posedness and gauge freedom as they are usually framed in the NR
literature.

\subsection{The 3+1 Split}\label{Sec:3+1}

Suppose that we are given a solution to Einstein's
equations~\eqref{eqn:EEs} expressed in coordinates~$x^\mu$ in a
coordinate patch. Then we may map from an abstract
tensor~$T_{a_1\dots a_k}{}^{b_1\dots b_l}$ to the
components~$T_{\mu_1\dots \mu_k}{}^{\nu_1\dots \nu_l}$ in the
coordinate basis defined by~$\p_\alpha^a$ and~$\nabla_ax^\mu$ simply
by contracting on every abstract index with elements of the basis,
\begin{align}
  T_{\mu_1\dots \mu_k}{}^{\nu_1\dots \nu_l}=T_{a_1\dots a_k}{}^{b_1\dots b_l}
  (\p_{\mu_1}^{a_1}\dots\p_{\mu_k}^{a_k})
  (\nabla_{b_1}x^{\nu_1}\dots\nabla_{b_l}x^{\nu_l})\,. 
\end{align}
Likewise the tensor itself can be reconstructed from the components
using
\begin{align}
  T_{a_1\dots a_k}{}^{b_1\dots b_l}=
  T_{\mu_1\dots \mu_k}{}^{\nu_1\dots \nu_l}
  (\nabla_{a_1}x^{\mu_1}\dots\nabla_{a_l}x^{\mu_l})
  (\p_{\nu_1}^{b_1}\dots\p_{\nu_k}^{b_k})
\,. 
\end{align}
We suppose now that~$x^\mu=(t,x^i)$ with level sets~$\Sigma_t$ of~$t$
spacelike, in the sense that vectors tangent to~$\Sigma_t$ are
spacelike. We then refer to~$t$ as a time coordinate. The collection
of level sets~$\Sigma_t$ is called a foliation of the
spacetime. Individual level sets are equivalently referred to as
either instants of time or slices. We define the lapse
function~$\alpha$ by,
\begin{align}
\alpha^{-2}=-g^{ab}\nabla_at\nabla_bt.
\end{align}
We may now define the future directed unit normal to slices of
constant~$t$ by
\begin{align}
n^a=-\alpha\nabla^at\,.
\end{align}
We define a projection operator through
\begin{align}
  \perp^a\!{}_b=\delta^a{}_b+n^an_b\,.
\end{align}
The lapse measures the proper time elapsed between neighboring slices
of the foliation. Direct computation reveals
that~$\perp^a\!{}_c\perp^c\!{}_b=\perp^a\!{}_b$, so that this indeed
defines a projection. The core idea of the~$3+1$ decomposition is that
we should never deal directly with general tensors. Instead, we
decompose all tensors into a set of objects that exist naturally
within the foliation. For example, given a general vector field~$v^a$,
we may define,
\begin{align}
 v^n\equiv v^a n_a \,,\qquad (\perp\!v)^a\equiv \perp^a\!{}_b v^b\,.
\end{align}
The vector field can be reconstructed from its normal~$v^n$ and
spatial~$(\perp\!v)^a$ components 
\begin{align}
 v^a = (\perp\!v)^a -n^a v^n \,.
\end{align}
The spatial part of the vector is orthogonal to the normal
vector~$(\perp\!v)^an_a=0$. The generalization of this decomposition
to higher rank tensors follows the obvious pattern. Any tensor which
is orthogonal to~$n^a$ on every index is called spatial. Since the
normal vector depends upon the time coordinate, this definition is
made with respect to a foliation. Consequently every spatial vector is
spacelike, but not vice-versa. Our task now is to apply the idea of
decomposition to the objects that appear in GR.

These objects may be naturally organized in a hierarchy according to
the number of derivatives that they (implicitly) contain. At zeroth
order we have the coordinates~$x^\mu=(t,x^i)$. Then at first order we
have the metric~$g_{ab}$, the coordinate basis
vectors~$\p_\alpha^a=(\p_t^a\equiv t^a,\p_i^a\equiv \p_{x^i}^a)$ and
the basis one-forms~$dx^\mu$ (which can also be
written~$\nabla_ax^\mu$). Recall that, by definition, we have the
relationships
\begin{align}
  \delta^a{}_b= \p_\alpha^a \nabla_bx^\alpha \,, \qquad
  \delta^\alpha{}_\beta= \p_\beta^a \nabla_ax^\alpha=\p_\beta x^\alpha\,,
\end{align}
which we use without further comment in the manipulations below. Next
are the connection
coefficients~${}^{\tiny{(4)}}\Gamma^\alpha{}_{\mu\nu}$, and at third
order, or equivalently with two derivatives of the metric, the
curvature~${}^{\tiny{(4)}}R_{abcd}$. Finally for our present needs,
the second Bianchi identities
\begin{align}
  \nabla_{[a}{}^{\tiny{(4)}}R_{bc]de}=0\,,
  \label{eqn:B2}
\end{align}
and any contractions thereof lie at the fourth level. We treat each
level of this hierarchy in turn.

\begin{figure}[t!]
  \vspace{0.3cm}
  \begin{center}
    \includegraphics[width=0.55\textwidth]{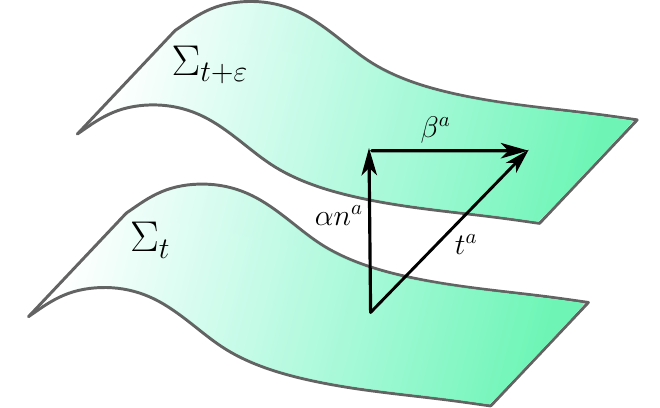}
  \end{center}
  \caption{Level sets~$\Sigma_t$ and neighbor~$\Sigma_{t+\varepsilon}$ of
    the foliation, together with a depiction of the relationship
    between the time vector~$t^a$, the lapse function~$\alpha$,
    the shift vector~$\beta^a$, and the future pointing unit
    normal to slices of the foliation~$n^a$.
    \label{fig:3+1_Foliation}
  }
\end{figure}

There is nothing do to treat the coordinates~$x^\mu=(t,x^i)$ as they
constitute a collection of scalar fields rather than a tensor.

Moving on the first level, the basis vectors decompose as,
\begin{align}
 t^an_a&=-\alpha \,,   &\perp\!t^a\equiv\beta^a\,,\nonumber\\
 \p_i^an_a&=0 \,,  &\perp\!\p_i^a=\p_i^a\,,\label{eqn:basis_decomposition}
\end{align}
where here and in what follows we adopt the convention that the
appearance of the projection operator without indices means projection
on all open abstract indices on the object directly to the right of
it. We have defined the shift vector~$\beta^a$ as the spatial part of
the time vector~$t^a$. It encodes the difference between the integral
curves of the time vector and the normal vector. Rearranging, we have
\begin{align}
  t^a=\alpha n^a + \beta^a\,,\qquad \p_i^a= \perp\!\p_i^a
  \label{eqn:basis_decomposition_reverse-form}
\end{align}
and, unsurprisingly, we see that the `spatial derivatives'~$\p_i$ are
spatial according to our definition without projection. The basis
one-forms~$dx^\mu$ are then given by,
\begin{align}
   n^a\nabla_at&= \alpha^{-1}\,,\qquad &\perp\!\nabla_at=0\,,\nonumber\\
  n^a\nabla_ax^i &= -\alpha^{-1}\beta^i \,,\qquad &\perp\!\nabla_ax^i
  = \gamma_a{}^b\nabla_ax^i=\gamma_a{}^i\,.\label{eqn:co-basis_decomposition}
\end{align}
These relationships may be rewritten as,
\begin{align}
  \nabla_at= -\alpha^{-1}n_a\,, \quad
  \nabla_ax^i = \alpha^{-1}\beta^i n_a + \perp\!\nabla_ax^i \,.
  \label{eqn:co-basis_decomposition_reverse-form}
\end{align}
The geometry of the foliation and the relationship between the normal
and time vectors is sketched in Fig.~\ref{fig:3+1_Foliation}. The
metric decomposes as
\begin{align}
  g_{nn}&=-1\,, \qquad \perp\!g_{an}=0\,, \qquad
  \gamma_{ab}=\perp\!g_{ab}= \perp_{ab} = g_{ab}+n_an_b \,.
  \label{eqn:metric_decomposition}
\end{align}
In the last expressions here we define the spatial metric, which may
be thought of geometrically as a metric on slices of the foliation
induced by~$g_{ab}$. Clearly the spatial metric is positive definite
when acting on spatial vectors~$\gamma_{ab}v^av^b\geq0$. Observe that
within the~$3+1$ split, the metric is special because it was used to
define the unit normal vector~$n^a$. Consequently
rearranging~\eqref{eqn:metric_decomposition}, we arrive at a
particularly simple expression
\begin{align}
  g_{ab}=\gamma_{ab}-n_an_b\,.
\end{align}
Nevertheless, observe that the decomposition of the metric itself used
the approach sketched above in a completely boiler-plate manner.

If we work in the coordinate basis~$\p_\alpha^a$, spatial tensors take
a particularly simple form. Consider once more an arbitrary spatial
vector~$v^a$. Expanding out the orthogonality condition~$v^an_a=0$, we
find,
\begin{align}
  0=v^an_a=-v^0\alpha\,,
\end{align}
which means that the time component of a spatial vector must
vanish~$v^0=0$ at any point where the foliation has non-vanishing
lapse. For an arbitrary spatial covector, we instead have
\begin{align}
  0=n^av_a= \alpha^{-1}v_0-\alpha^{-1}\beta^iv_i \,,
\end{align}
which means that the time component is given by the inner product with
the shift vector, which is itself spatial, so that so long as the
lapse is bounded we have,
\begin{align}
  v_0=\beta^iv_i\,.
\end{align}
The generalization of this result to tensors of arbitrary valence is
straightforward. We conclude that, in a regular foliation with finite
non-vanishing lapse, if we know the shift vector, then the spatial
components of an arbitrary spatial tensor suffice to reconstruct the
full tensor. We will use this to express the full Einstein equations
in our coordinate basis below after the remainder of the geometric
decomposition.

In the second level of the hierarchy we need to obtain a complete
representation of the connection~$\nabla_a$. For the reasons discussed
in the last paragraph, we need only know how to compute derivatives of
spatial tensors within the~$3+1$ formalism. Given an arbitrary spatial
tensor we define the spatial covariant derivative by total projection
of the spacetime covariant derivative
\begin{align}
  D_aT_{a_1\dots a_k}{}^{b_1\dots b_l}
  = \perp\!\nabla_a T_{a_1\dots a_k}{}^{b_1\dots b_l}\,.
\end{align}
It turns out that the spatial covariant derivative is nothing more
than the Levi-Civita derivative of the spatial metric. This can be
verified by verifying first that it is torsion free
\begin{align}
  2 D_{[a}D_{b]}\phi = 2\perp\!\nabla_{[a}\left(\nabla_{b]}\phi
  +n_{b]}\Lie_n\phi\right)
  = 2\perp\!\nabla_{[a}n_{b]} \Lie_n\phi = 0\,.\label{eqn:D_torsion_free}
\end{align}
Here the last term vanishes by Frobenius theorem because~$n^a$ is
hypersurface orthogonal, square parentheses~$[\,\,\,\,]$ on indices
denote antisymmetrization with the usual normalization
of~\cite{Wal84}, and we employ the standard notation for the
Lie-derivative. Second, we see that it is compatible with the spatial
metric,
\begin{align}
  D_a\gamma_{bc}\equiv \perp\!\nabla_a(g_{bc}+n_bn_c)=0\,.
\end{align}
Observe that with this definition the spatial coordinate basis
from~\eqref{eqn:co-basis_decomposition_reverse-form} can be rewritten
as~$\nabla_ax^i = \alpha^{-1}\beta^i n_a+D_ax^i$. To capture pieces of
the projected covariant
derivative~$\perp\!^c{}_a\nabla_cT_{a_1\dots a_k}{}^{b_1\dots b_l}$
normal to the level set~$\Sigma_t$ we need a representation of
derivatives of~$n_a$. For this we define the acceleration vector~$a_a$
and the extrinsic curvature~$K_{ab}$, which are given, respectively,
by
\begin{align}
  a_a=n^b\nabla_bn_a\,, \qquad
  K_{ab}=-\gamma^c{}_a\nabla_cn_b\,.
\end{align}
Both are spatial by virtue of the unit normalization of~$n^a$. One
should be aware that in the mathematics literature the extrinsic
curvature is occasionally defined with the opposite sign convention.
As in Eq.~\eqref{eqn:D_torsion_free}, hypersurface orthogonality
of~$n_a$ simplifies matters, allowing us to
rewrite~$a_a=D_a\ln\alpha$. The covariant derivative of~$n_a$ is
completely represented through these variables and can now be
expressed as
\begin{align}
  \nabla_an_b=-K_{ab}-n_aa_b\,.
\end{align}
The same argument given below equation~\eqref{eqn:D_torsion_free}
shows that the extrinsic curvature is symmetric. Using orthogonality
of~$n^a$ with spatial~$T_{a_1\dots a_k}{}^{b_1\dots b_l}$ it follows
that
\begin{align}
  n^{a_1}\perp\!^c{}_a\nabla_cT_{a_1\dots a_k}{}^{b_1\dots b_l}
  =-(\perp\!^c{}_a\nabla_cn^{a_1})T_{a_1\dots a_k}{}^{b_1\dots b_l}
  =K_a{}^{a_1}T_{a_1\dots a_k}{}^{b_1\dots b_l}\,,\label{eqn:n-dot-perp-nabla-T}
\end{align}
and likewise with the contraction with~$n^a$ on any slot within the
derivative, so that we now have a complete representation of any
covariant derivative of a spatial tensor projected with~$\perp\!$ on
the derivative index itself. The extrinsic curvature plays yet another
role, however, as it can be viewed morally as a `time derivative' of
the spatial metric, through the relation
\begin{align}
  K_{ab}=-\tfrac{1}{2}\Lie_n\gamma_{ab}\,.\label{eqn:gamma_evolve}
\end{align}
There is variability in the literature in the representation of the
`time derivative' of an arbitrary spatial tensor under the~$3+1$ split
that depends on the specific equations under consideration, but this
serves as the canonical example. The core idea is always to introduce
the Lie derivative along a vector field with non-vanishing contraction
with~$n^a$ as a dependent variable, and then project the resulting
object on every remaining open index. For instance
for~$T_{a_1\dots a_k}{}^{b_1\dots b_l}$, we might introduce
\begin{align}
S_{a_1\dots a_k}{}^{b_1\dots b_l}=\perp\!\Lie_nT_{a_1\dots a_k}{}^{b_1\dots b_l}\,.
\end{align}
The remainder of the Lie-derivative
of~$T_{a_1\dots a_k}{}^{b_1\dots b_l}$ along~$n^a$ is given by
relationships akin to equation~\eqref{eqn:n-dot-perp-nabla-T} but with
the acceleration~$a_a$ playing the role of the extrinsic
curvature. Again contracting on the first index of a given type as a
representative example, we have,
\begin{align}
  n^{a_1}\Lie_nT_{a_1\dots a_k}{}^{b_1\dots b_l}=0\,,
\end{align}
when the contraction is made with any index downstairs, and
\begin{align}
  n_{b_1}\Lie_nT_{a_1\dots a_k}{}^{b_1\dots b_l}
  =-a_{b_1}T_{a_1\dots a_k}{}^{b_1\dots b_l}\,.
\end{align}
otherwise. In fact we have now introduced enough variables to achieve
a complete description of the connection coefficients in our
coordinate basis. To verify this we simply use the general definition,
\begin{align}
  {}^{\tiny (4)}\Gamma^\alpha{}_{\mu\nu} =
   \p_{\mu}^a\,(\nabla_a \p_\nu^b)\,\nabla_bx^\alpha\,,
\end{align}
express our basis vectors as
in~\eqref{eqn:basis_decomposition_reverse-form},
and~\eqref{eqn:co-basis_decomposition_reverse-form}, and collect all
possible contractions and projections, which results in
\begin{align}
  {}^{\tiny (4)}\Gamma^n{}_{nn}&=-\p_n\ln\alpha\,,\nonumber\\
  {}^{\tiny (4)}\Gamma^n{}{}_{in}&=-a_i=-\p_i\ln\alpha\,,\nonumber\\
  {}^{\tiny (4)}\Gamma^{\alpha}{}_{nn}\perp\!^i{}_{\alpha}&
  =\alpha^{-1}\p_n\beta^i+a^i\,,\nonumber\\
  {}^{\tiny (4)}\Gamma^{\alpha}{}_{in}\perp\!^j{}_{\alpha}&
  =\alpha^{-1}\p_i\beta^j-K_i{}^j\,,\nonumber\\
  {}^{\tiny (4)}\Gamma^{n}{}_{ij}&=K_{ij}\,,\nonumber\\
  {}^{\tiny (4)}\Gamma^{\alpha}{}_{ij}\perp\!^k{}_{\alpha}
  &= \p_i^a (D_a \p_j^b) D_bx^k \equiv \Gamma^{k}{}_{ij}\,.
  \label{eqn:connection_decomposition}
\end{align}
Observe that, despite the name, the extrinsic curvature really belongs
to the connection. We realized already above that the spatial
covariant derivative is the Levi-Civita covariant derivative of the
spatial metric. It follows that the spatial connection
coefficients~$\Gamma^k{}_{ij}$ are just the associated Christoffel
symbols,
\begin{align}
  \Gamma^k{}_{ij}=\tfrac{1}{2}\gamma^{kl}(\p_i\gamma_{jl}+\p_j\gamma_{il}
  -\p_l\gamma_{ij})\,.
\end{align}
In a coordinate basis the spacetime connection
coefficients~${}^{\tiny (4)}\Gamma^{\alpha}{}_{\mu\nu}$ have~$40$
algebraically independent components. Taking into account the symmetry
of~$K_{ij}$ and~$\Gamma^k{}_{ij}$ in the downstairs indices and adding
the number of components of each line
of~\eqref{eqn:connection_decomposition}, we have~$1+3+3+9+6+18=40$,
confirming that we have captured the spacetime object completely.

We now move on to the third level in our derivative hierarchy, the
geometric decomposition of curvature quantities. In~$3+1$ dimensions
there are~$20$ algebraically independent components of the spacetime
Riemann curvature~${}^{\tiny{(4)}}R_{abcd}$, and we therefore
seek~$20$ algebraically independent equations upon decomposition. In
other words; symmetries of the Riemann tensor vastly reduce the number
of different contractions and projections we must consider. We start
by projecting on every index, which reveals,
\begin{align}
  \perp {}^{\tiny{(4)}}R_{abcd} &= R_{abcd}+2K_{a[c}K_{d]b}\,.
                                  \label{eqn:Gauss_eqn}
\end{align}
Here we have introduced the spatial Riemann tensor~$R_{abcd}$, the
curvature associated with the spatial covariant derivative, defined as
usual through commutation,
\begin{align}
  2D_{[a}D_{b]}w_c &= R_{abc}{}^dw_d\,,
\end{align}
with~$w_a$ an arbitrary spatial
one-form. Equation~\eqref{eqn:Gauss_eqn} is known as the Gauss
equation. It is straightforwardly derived by
contracting~${}^{\tiny{(4)}}R_{abcd}$ on the last index with an
arbitrary spatial vector, projecting on the remaining indices and
using the definition of the spatial covariant derivative~$D_a$ and
extrinsic curvature~$K_{ab}$. Recall that, although we are using
abstract spacetime indices, since we work with spatial tensors we need
only count their spatial components. In~$3$ dimensions the Riemann
tensor has~$6$ algebraically independent components. A useful
expression of this fact is that in the~$3+1$ setting the spatial
Riemann tensor~$R_{abcd}$ is determined entirely by the spatial Ricci
tensor~$R_{ab}=R_{acb}{}^c$ and its trace, the spatial Ricci
scalar~$R=\gamma^{ab}R_{ab}$,
\begin{align}
  R_{abcd} = 2 \left(\gamma_{a[c} R_{d]b}- \gamma_{b[c} R_{d]a}\right)
  - R \gamma_{a[c}\gamma_{d]b}\,,\label{eqn:3d_No_Weyl}
\end{align}
which can also be viewed as the fact that in~$3$ dimensions the Weyl
tensor vanishes identically. Next we have the Codazzi-Mainardi
equations, which are built by contracting once with the normal
vector~$n^a$ before projecting on the remaining indices, giving
\begin{align}
  \perp\, {}^{\tiny{(4)}}R_{abcd}n^d &= - 2 D_{[a} K_{b]c} \,.
  \label{eqn:Codazzi-Mainardi}
\end{align}
The derivation of this result follows the exact same pattern sketched
above for the Gauss equation. Counting here is a little more
subtle. For each of the~$3$ independent components of the index
without symmetrization, there are~$3$ equations associated with the
antisymmetrized indices, giving a total of~$9$ equations. But the
totally antisymmetric part also vanishes by the first spacetime
Bianchi identity~${}^{\tiny{(4)}}R_{[abc]d}=0$. In~$3$ dimensions this
amounts to one more equation that we must subtract, giving~$8$
algebraically independent components in total
in~\eqref{eqn:Codazzi-Mainardi}. The same combination of contractions
and projections can in fact be used together with the Bianchi identity
to efficiently obtain an evolution equation for the spatial connection
coefficients, but this not needed presently. Finally, the Ricci
equations are obtained by contracting twice with the normal vector and
projecting on the remaining indices, which gives
\begin{align}
  \perp\, {}^{\tiny{(4)}}R_{abcd}n^b n^d
  &= \Lie_n K_{ab}+\alpha^{-1}D_aD_b\alpha + K^c{}_aK_{bc}\,,
    \label{eqn:Ricci_eqn}
\end{align}
again by following the procedure sketched above. The symmetries of the
Riemann tensor actually make the projection superfluous, but
maintaining it makes counting more transparent. Since we have now a
symmetric spatial tensor, there are~$6$ algebraically independent
components here. Double-checking, from~\eqref{eqn:Gauss_eqn},
\eqref{eqn:Codazzi-Mainardi} and~\eqref{eqn:Ricci_eqn} we have a total
of~$6+8+6=20$ algebraically independent components, as required.

Moving now to the fourth level of the hierarchy, the highest that we
require here, the decomposition of the second Bianchi
identities~\eqref{eqn:B2} proves useful. These correspond to
algebraically~$20$ independent equations, with four independent
contractions of interest. We begin by projecting on every index, which
reveals the second Bianchi identity for~$D_a$,
\begin{align}
  0=\perp\! \nabla_{[a}{}^{\tiny{(4)}}R_{bc]de} = D_{[a} R_{bc]de} \,,
  \label{eqn:D_B2}
\end{align}
which could also be viewed as a consequence of the fact that the
spatial covariant derivative is itself torsion-free, again as a result
of the fact that~$n^a$ is hypersurface orthogonal. These correspond
to~$3$ equations, consistent with the number of contracted Bianchi
identities in~$3$ dimensions, which can be seen by substitution
of~\eqref{eqn:3d_No_Weyl} into~\eqref{eqn:D_B2}. Next we contract with
the normal vector on one of the non-explicitly antisymmetrized indices
and project the rest. This gives
\begin{align}
  0 = - \perp\! \nabla_{[a}{}^{\tiny{(4)}}R_{bc]de} n^e
  = 2D_{[a}D_{b}K_{c]d} - K^e{}_{[a}R_{bc]de} \,,
\end{align}
which contains another~$3$ algebraically independent equations. Viewed
from the perspective of the spatial covariant derivative, this
identity holds simply by the definition of the spatial Riemann
curvature and the Bianchi identity~\eqref{eqn:D_B2}. Contracting
instead with the normal vector on the antisymmetrized indices and
projecting, we have
\begin{align}
  0 &= 3\perp\! \nabla_{[a}{}^{\tiny{(4)}}R_{bc]de} n^a\nonumber\\
    &= \perp\!\Lie_n\big(\!\perp\!{}^{\tiny{(4)}}R_{bcde}\big)
      -2\!\perp\!{}^{\tiny{(4)}}R_{bca[d} K_{e]}{}^a
      +2\gamma^{fg}\!\perp\!{}^{\tiny{(4)}}R_{feg[b} K_{c]d}
      -2\gamma^{fg}\!\perp\!{}^{\tiny{(4)}}R_{fdg[b} K_{c]e}\nonumber\\
    & -2D_{b}\big(D_{[d}K_{e]c}\big)
      +2D_{c}\big(D_{[d}K_{e]b}\big)
      +4\big(D_{[d}K_{e][b}\big)a_{c]} 
      +4\big(D_{[b}K_{c][d}\big)a_{e]}\nonumber\\
    & +2\perp\!{}^{\tiny{(4)}}R_{d[b} K_{c]e}
      -2\perp\!{}^{\tiny{(4)}}R_{e[b} K_{c]d}\,,
  \label{eqn:B2_decomposed_R}
\end{align}
which encodes~$6$ equations. In this equation, for brevity, we have
not expanded out~$\perp\!{}^{\tiny{(4)}}R_{abcd}$
from~\eqref{eqn:Gauss_eqn}. Finally, we contract once with the unit
normal in each of the two blocks of indices,
\begin{align}
  0 &= 3\perp\! \nabla_{[a}{}^{\tiny{(4)}}R_{bc]de}  n^a n^e\nonumber\\
    &= -2\perp\! \Lie_n(D_{[b}K_{c]}{}^e)\gamma_{de}
      +2KD_{[b}K_{c]d} + K_{c}{}^e\left(2D_{[d}K_{e]b}-D_bK_{de}
      +D_b(\gamma_{de} K)\right) \nonumber\\
    &-2K_b{}^eD_{[d}K_{e]c}+2D_{[b}R_{c]d}+K_{eb}D_cK_{de}-K_{bd}D_cK
      +2a_{[b}\left(R_{c]d}+K_{c]d}K-K_{c]}{}^eK_{de}\right)\nonumber\\
    &-a^e\left(R_{bcde}+2K_{b[d}K_{e]c}\right)
      -2^{-1}D_{[b}(\perp\!{}^{\tiny{(4)}}R_{c]d})
      -2a_{[b}\!\perp\!{}^{\tiny{(4)}}R_{c]d}\,,
      \label{eqn:B2_decomposed_K}
\end{align}
which results in the remaining~$8$ algebraically independent
equations.

To this point our discussion was entirely geometrical. Given a
spacetime manifold~$M$, Lorentzian metric~$g_{ab}$ with an associated
Levi-Civita connection~$\nabla_a$, coordinates~$x^\mu=(t,x^i)$ whose
time coordinate~$t$ has spacelike level sets and associated coordinate
basis vectors~$\p_\alpha$, and one-forms~$dx^\mu$, we have understood
how to rewrite the geometry in a~$3+1$ decomposed form. We now use
this formalism to reformulate Einstein's
equations~\eqref{eqn:EEs}. For this we could apply our recipe above in
a purely agnostic fashion to the field equations. But to streamline
the derivation it is convenient to begin with the observation that
in~$3+1$ dimensions the Einstein tensor can be decomposed in the form
\begin{align}
  G_{ab}=  G_{nn} (n_an_b+\gamma_{ab}) + 2 n_{(a}\perp\!G_{b)n}
  + \perp\!{}^{\tiny{(4)}}R_{ab}
  -\gamma_{ab}{}^{\tiny{(4)}}R_{cd}\gamma^{cd}\,,
\end{align}
where round parentheses~$(\,\,\,\,)$ on indices denote symmetrization
with the usual normalization of~\cite{Wal84}. Thus we may consider the
three pieces~$G_{nn}$, $\perp\!G_{an}$
and~$\perp\!{}^{\tiny{(4)}}R_{ab}$. We define variables for the
decomposition of the stress-energy tensor, namely
\begin{align}
  \rho=T_{ab}n^an^b\,,
  \quad j_a= - \perp\! T_{ab}n^b\,,
  \quad s_{ab} = \perp\! T_{ab}\,.
\end{align}
the energy density, momentum density and spatial stress tensors,
respectively, as viewed by normal observers who travels along integral
curves of~$n^a$. Straightforward application of the~$3+1$
decomposition on~$G_{nn}$ then reveals the first of the field
equations as
\begin{align}
  H=2G_{nn}-16\pi T_{nn}=R + K^2 - K^{ab}K_{ab}- 16 \pi \rho=0\,.
  \label{eqn:Ham}
\end{align}
Observe that no explicit time derivatives occur here (in this
formalism these would occur in the form~$\Lie_n$). For this reason,
this equation constraints the permissible data at any instant of
time~$\Sigma_t$. It is called the Hamiltonian constraint. Taking now
the~$\perp\!G_{an}$ and using a trace of the Codazzi-Mainardi
formula~\eqref{eqn:Codazzi-Mainardi} we obtain,
\begin{align}
  M_a= -\perp\!G_{an} = D^bK_{ab}-D_aK - 8 \pi j_a = 0\,.\label{eqn:Mom}
\end{align}
Again, there are no explicit time derivatives and so these equations
restrict permissible data at all times. This is known as the momentum
constraint. For the last piece~$\perp\!{}^{\tiny{(4)}}R_{ab}$, we
project the trace-reversed Einstein equations, obtaining
\begin{align}
  \Lie_nK_{ab} &= -\alpha^{-1}D_aD_b\alpha + R_{ab}
                 -2K^c{}_aK_{bc}+KK_{ab}\nonumber\\
               &\quad + 8 \pi \left[ s_{ab}
                 - \tfrac{1}{2}\gamma_{ab}s
                 + \tfrac{1}{2}\gamma_{ab}\rho\right]\,,
\end{align}
These are now true evolution equations which, in some sense, encode
the dynamical behavior of GR. 

It remains to offer a little tidying up. Observe that the Einstein
equations do not determine the time evolution of~$\gamma_{ab}$ but
rather that of the extrinsic curvature. This is no problem, however,
since~\eqref{eqn:gamma_evolve} already told us that the extrinsic
curvature can be viewed as the time derivative of the metric. For
practical numerical purposes we need to express the field equations
not as Lie derivatives of abstract index tensors but directly with
partial derivatives of their components. This is easily achieved by
using the relation~\eqref{eqn:basis_decomposition_reverse-form} to
replace the normal vector with the time vector~$t^a$ and expanding out
the equations in our coordinate basis. Doing so, we finally arrive at
the Arnowitt-Deser-Misner (ADM), or York, evolution equations,
\begin{align}
\p_t\gamma_{ij}&=-2\alpha K_{ij}+\Lie_\beta\gamma_{ij}\,,\nonumber\\
\p_tK_{ij}&=-D_iD_j\alpha
            +\alpha\left[R_{ij}-2K^k{}_iK_{jk}+KK_{ij}\right]\nonumber\\
          &\quad + 4 \pi \alpha\left[ \gamma_{ij}(s-\rho) - 2s_{ij}\right]
            +\Lie_\beta K_{ij}\,.
            \label{eqn:ADM}
\end{align}
We make the usual abuse of notation when expressing derivatives, in a
basis, writing
\begin{align}
  D_iD_j\alpha\equiv (\p_i^a\p_j^b) D_aD_b\alpha\,,
  \quad \Lie_\beta \gamma_{ij} \equiv (\p_i^a\p_j^b) \Lie_\beta \gamma_{ab}\,,
\end{align}
and so forth. If we take a divergence of the Einstein equations, the
left-hand side vanishes identically, giving~$\nabla^bT_{ab}=0$, which
can be interpreted as the local conservation of energy and
momentum. Under the~$3+1$ decomposition these four equations can be
written as
\begin{align}
  \p_t\rho&=\alpha[D^ij_i+2a^ij_i-K\rho-K^{ij}s_{ij}]
            +\Lie_\beta\rho\,,\nonumber\\
  \p_tj_i&=-\alpha[D^js_{ij}+a^js_{ij}-Kj_i+a_i\rho]
           +\Lie_\beta j_i\,.\label{eqn:Local_Cons}
\end{align}
Observe that the lapse and shift are completely absent from the
constraints, and furthermore that they have no equations of
motion. Instead, they encode the freedom in the choice of
coordinates. In NR this is usually viewed as the gauge freedom of the
theory. We asserted above that the Hamiltonian~\eqref{eqn:Ham} and
momentum~\eqref{eqn:Mom} constraints restrict the permissible data at
each instant of time. This can only be self-consistent if the
evolution equations~\eqref{eqn:ADM} respect the constraints, in the
sense that the time development of constraint-satisfying initial data
remains constraint-satisfying. It turns out that, under the evolution
equations~\eqref{eqn:ADM}, the constraints evolve according to the
formal evolution system
\begin{align}
  \p_tH&= -2\alpha D^iM_i - 4M^iD_i\alpha + 2\alpha K H + \Lie_\beta H
         \,,\nonumber\\
  \p_tM_i&= -\tfrac{1}{2}\alpha D_iH +\alpha K M_i
           - (D_i\alpha) H + \Lie_\beta M_i  \,.
           \label{eqn:ADM_constraint_subsystem}
\end{align}
These equations can be obtained, for instance, by taking suitable
traces of~\eqref{eqn:B2_decomposed_R} and~\eqref{eqn:B2_decomposed_K},
plugging in the Einstein equations and using~\eqref{eqn:Local_Cons},
or equivalently by~$3+1$ decomposition of the spacetime contracted
Bianchi identities. Obviously~$H=M_a=0$ is a solution to these
equations. Therefore given constraint satisfying initial data that can
be evolved in time, there ought to be at least one solution that is
constraint satisfying.

Putting aside for now any subtleties associated with the stress-energy
tensor, we can now give a complete geometric formulation of the
initial value (Cauchy) problem for GR.  At an instant of
time~$\Sigma_t$ we choose~$(\gamma_{ab},K_{ab})$ satisfying the
Hamiltonian and momentum constraints, which then evolve according to
the evolution system~\eqref{eqn:ADM}. This leads to the concept of
free-evolution in NR, in which the idea is to solve the constraints
just once for initial data and then evolve those data in time, relying
on closure of the constraint subsystem to guarantee they are satisfied
at later times. In practice numerical error violates the constraints,
but these violations should converge away with increasing
resolution. Strategies to obtain initial data of interest are
discussed in the subsection~\ref{Sec:ID} next. To be sure that initial
data give rise to a time development that can be reliably calculated
by numerical methods, we must understand whether or not the evolution
PDE we are solving is well-posed. Well-posedness is the requirement
that a PDE problem admits a unique solution that depends continuously
on the given data. This is discussed for evolution problems in
subsection~\ref{Sec:WP}. The choice of gauge has a direct effect on
well-posedness, and so is fundamental in the construction of the
free-evolution formulations of GR. This freedom is discussed below in
subsection~\ref{Sec:Free-Evolution}.

\subsection{Initial Data}
\label{Sec:ID}

Our aim in the construction of initial data is to find a configuration
of the variables that models a physical scenario of interest whilst
simultaneously solving the Hamiltonian and momentum constraints. To
fulfill the former condition it is often advantageous to append
additional equations to the constraints to impose physical
restrictions on the data. To keep the discussion here brief, we review
only the most common approach.

As we saw above, the Hamiltonian and momentum constraints
\begin{align}
R + K^2 - K^{ab}K_{ab}&=16 \pi \rho\,,\nonumber\\
D^bK_{ab}-D_aK &= 8 \pi j_a\,,
\end{align}
restrict the permissible data at each instant of time~$\Sigma_t$. They
are a collection of four equations for the twelve components of the
spatial metric, extrinsic curvature and four components of the
stress-energy tensor. Since this system of equations is highly
under-determined and does not have a fixed PDE character, many
different approaches may be taken to find solutions. What all
strategies have in common is that decisions must be made about what
part of the initial data will be specified, what `dependent variables'
will be introduced and solved for, what data will be given to those
variables, and where it should be given.

In asymptotically flat spacetimes, the class most prevalent in NR, the
metric and extrinsic curvature reduce to the flat-metric far away from
a gravitating source. An attractive approach is thus to specify data
for the dependent variables at spatial infinity, where they ought to
be simple. In practice this means that we would like to solve a
boundary value problem. The canonical PDE with a well-posed boundary
value problem is the Laplace equation, which is of elliptic type. For
these reasons it is popular to formulate the constraints as a system
of second order elliptic PDEs and solve a boundary value
problem. There are various different approaches even to this, but we
shall focus on the archetype, the conformal transverse-traceless
decomposition.

We begin by making a conformal decomposition of the metric,
\begin{align}
  \gamma_{ij} &= \psi^4\tilde{\gamma}_{ij}\,.
\end{align}
We use the common convention to raise and lower the indices of objects
with a tilde using the inverse conformal
metric~$\gamma^{ij} = \psi^{-4}\tilde{\gamma}^{ij}$. We denote
by~$\tilde{D}_i$ the Levi-Civita connection
of~$\tilde{\gamma}_{ij}$, write the associated curvature
as~$\tilde{R}_{ijkl}$, and denote contractions
similarly. Straightforward calculations relate the spatial and
conformal curvature scalars
\begin{align}
  R &= \psi^{-4} \tilde{R} - 8 \psi^{-5}\tilde{\Delta}\psi \,,
\end{align}
where the conformal Laplace operator
is~$\tilde{\Delta}\equiv\tilde{D}^i\tilde{D}_i$. Plugging this into
the Hamiltonian constraint, we obtain an elliptic equation for the
dependent variable~$\psi$. Besides the fact that the conformal
decomposition naturally introduces an elliptic operator on the
conformal factor, it is aesthetically pleasing that in this
construction no single metric component is singled out for special
treatment.

The momentum constraint contains only one derivative of the extrinsic
curvature, and so to render it a second order equation the obvious
maneuver is to introduce a potential. This is the transverse-traceless
part of the setup. We write the extrinsic curvature as
\begin{align}
  K^{ij} &= \psi^{-10}\tilde{A}^{ij}+\tfrac{1}{3}\gamma^{ij}K\,,
\end{align}
and place the ansatz
\begin{align}
  \tilde{A}^{ij} &= \tilde{D}^i\tilde{v}^j+\tilde{D}^j\tilde{v}^i
                   -\tfrac{2}{3}\tilde{\gamma}^{ij}\tilde{D}_k\tilde{v}^k
                   +\tilde{M}^{ij}\,,
\end{align}
on the conformal trace-free part~$\tilde{A}_{ij}$,
with~$\tilde{M}^{ij}$ an arbitrary symmetric trace-free
tensor. Plugging this into the momentum constraint, we again obtain a
second order elliptic equation, this time for the vector
potential~$\tilde{v}^i$, albeit with a more complicated elliptic
operator
\begin{align}
  \tilde{\Delta}_{\bf v}\tilde{v}^i=
  \tilde{\Delta}\tilde v^i+\tfrac{1}{3}\tilde{D}^i(\tilde{D}_j\tilde{v}^j)
  +\tilde{R}^i{}_jv^j\,,
\end{align}
than for the conformal factor.

Putting all of this together, the conformal transverse-traceless
decomposition gives the constraints the form
\begin{align}
  8 \tilde{\Delta}\psi - \tilde{R}\psi+\psi^{-7}\tilde{A}_{ij}\tilde{A}^{ij}
  -\tfrac{2}{3}\psi^5K^2+16\pi\psi^5\rho &=0\,,\nonumber\\
  \tilde{\Delta}_{\bf v}\tilde{v}^i + \tilde{D}_j\tilde{M}^{ij}
  - \tfrac{2}{3} \psi^6 \tilde{D}^iK - 8\pi\psi^{10} j^i &=0\,.
  \label{eqn:Constraints_ctt}
\end{align}
We choose data for the conformal metric~$\tilde{\gamma}_{ij}$, the
trace of the extrinsic curvature~$K$, the symmetric trace-free
tensor~$\tilde{M}^{ij}$ and pick a matter distribution, which will
generate~$\rho$ and~$j^i$. As mentioned above, in the routine
asymptotically flat setting we impose boundary conditions~$\psi=1$
and~$v^i=0$ at spatial infinity and then solve for~$(\psi,v^i)$
everywhere on~$\Sigma_t$. Yet another beautiful property of elliptic
equations is that existence theorems, when they can be applied,
guarantee that the solution exists on the entire domain. 

Once we have a concrete formulation of the constraints we can turn our
attention to making a choice of data that represents the scenario we
are interested in studying. In doing so there is frequently a
trade-off between the mathematical simplicity and accurately capturing
this scenario. For instance, working in vacuum~$\rho=j_i=0$, taking
the conformal metric flat~$\tilde{\gamma}_{ij}=\delta_{ij}$ and
choosing~$K_{ij}=0$, the Hamiltonian constraint reduces to the
flat-space Laplace equation and the momentum constraint is trivially
fulfilled. The Hamiltonian constraint is then easily solved. This
choice of initial data, called Brill-Lindquist data, represents a
number of superposed black holes instantaneously at rest. These are
evidently modeled after the Schwarzschild solution in isotropic
coordinates, which we recall has spatial metric
\begin{align}
  \gamma_{ij}dx^idx^j=\psi^4(dx^2+dy^2+dz^2)=
  \left(1+\tfrac{M}{2r}\right)^4(dx^2+dy^2+dz^2)\,,
  \label{eqn:SS_isotropic}
\end{align}
with~$r^2=x^2+y^2+z^2$ as usual. Since we are most often interested in
black holes in orbit around each other, this is too
restrictive. Working in vacuum with the same conformally flat metric,
choosing~$K=0$, called maximal slicing, and~$\tilde{M}^{ij}=0$, the
momentum constraint decouples from the conformal factor and admits an
exact solution, called the Bowen-York solution. This can be used to
give linear momentum and spin to the black holes, with the price that
the Hamiltonian constraint remains a nonlinear PDE. To solve it we
turn then to numerical methods. These data are called
`puncture-initial data'. They are among the most used binary black
hole configurations. Thus, Schwarzschild written in isotropic
coordinates serves as the prototype for one of the two main classes of
black hole initial data. Again the assumptions placed to simplify the
constraints are a direct shortcoming, since neither instants of time
in the Schwarzschild spacetime that carry linear momentum nor time
slices of the Kerr spacetime can be conformally flat. The prototype
for the second class of black hole initial data, called excision data,
is the Schwarzschild solution written in horizon penetrating
coordinates. The basic idea for the time-evolution of this data is
that, since the interior of the black hole can not effect what happens
outside, we may cut the black hole region out of the computational
domain. See below for a short discussion of what this entails. For
excision initial data it is more common to drop the assumptions of
conformal flatness and maximal slicing. Likewise when using more
sophisticated physical input such as that from Post-Newtonian
theory~\cite{PoiWil14}, simplifying mathematical assumptions must be
abandoned. In that case, the system~\eqref{eqn:Constraints_ctt}
remains coupled and so requires a general purpose numerical solver for
elliptic systems. For an up-to-date review of what is known rigorously
for this complete system, see~\cite{Car21}.

Setting up initial data for systems including fluid matter poses its
own complications. Initial data that are co-rotational or irrotational
can be constructed using the clean procedure as presented
in~\cite{BauSha10}. But for generic binary configurations it is common
to boost numerically constructed Tolman–Oppenheimer–Volkoff (TOV)
configurations to help construct the given data for whatever
formulation of the constraints (most often the
extended-conformal-thin-sandwich system) is being used. Another
subtlety lies at the treatment of the stellar surface, where the Euler
equations become formally singular. Further discussion of this can be
found in the review article~\cite{Tic16}.

\subsection{Well-posedness of time evolution PDEs}
\label{Sec:WP}

The complete specification of a PDE problem consists of the domain on
which the problem is to be solved, the data that is given and of
course the PDEs themselves. If a PDE problem admits a unique solution
that depends continuously on the given data in some norm we say that
the problem is well-posed (with respect to that norm). Well-posedness
is a fundamental requirement on a PDE problem, and is of practical
concern for numerical approximation. If a PDE is ill-posed, that is,
if there exists no norm in which it is well-posed, then solutions of
any approximation scheme will fail to converge to the continuum
solution (if it even exists) in the limit of infinite resolution.

In this subsection we state without proof the most relevant
well-posedness results for the initial value problem and initial
boundary value problem. For proofs, together with a numerical
perspective, we recommend the classic
textbooks~\cite{KreLor89,GusKreOli95}. For a discussion closest to the
language and needs of NR, we suggest~\cite{SarTig12,Hil13}, the
fascinating review~\cite{FriRen00}, and for a pure mathematical
treatment~\cite{Rin09b}. For an introduction to the long-term behavior
of nonlinear wave equations, we also recommend~\cite{Sog95}.

Consider the first order linear constant coefficient evolution system
\begin{align}
  \p_t\mathbf{u}&= \mathbf{A}^p\p_p\mathbf{u}+\mathbf{s}\,.
  \label{eqn:FT1S_CC}
\end{align}
The principal part of the system consists just of the terms involving
the highest derivatives. We assume that the components of the state
vector~$\mathbf{u}$, the principal part matrices~$\mathbf{A}^p$ and
the source terms~$\mathbf{s}$ are real. Observe that the source terms
could be of the
form~$\mathbf{s}=\mathbf{B}\mathbf{u}+\mathbf{s}'$. The initial value
problem consists of setting data,
\begin{align}
 \mathbf{u}(0,x)&= \mathbf{f}(x) \,,
\end{align}
and requesting a solution at times~$t>0$. In this context,
well-posedness of the initial value problem in a norm~$||\cdot||$ can
be characterized by the existence of positive constants~$\alpha,K$
such that an estimate
\begin{align}
||\mathbf{u}(t,\cdot)||\leq K e^{\alpha t}||\mathbf{f}(\cdot)||
\label{eqn:WP_CC_Energy_Estimate}
\end{align}
holds regardless of the choice of initial data.

Let~$s_p$ denote an arbitrary unit spatial covector. The principal
symbol~$\mathbf{P}^s$ in the~$s_p$ direction is defined
by~$\mathbf{P}^s=\mathbf{A}^s\equiv\mathbf{A}^ps_p$. The
system~\eqref{eqn:FT1S_CC} is called weakly hyperbolic if, for every
such covector, the principal symbol has real eigenvalues. These can be
interpreted as the speeds of propagation of the system. If,
furthermore, for every~$s_p$ the principal symbol has a complete set
of eigenvectors and the bound
\begin{align}
  |\mathbf{T}_s|+|\mathbf{T}_s^{-1}|\leq K\,,
  \label{eqn:SH_uniformity_condition}
\end{align}
holds uniformly in the direction~$s_p$ for some~$K>0$, then we say
that the system is strongly hyperbolic. Here~$\mathbf{T}_s$ is the
matrix that takes the right eigenvectors as columns and
where~$|\cdot|$ denotes here the spectral norm of a matrix. If there
exists a symmetrizer~$\mathbf{H}$ for the system, a symmetric positive
definite matrix such that~$\mathbf{H}\mathbf{A}^p$ is symmetric for
each~$p$, we say that the system is symmetric hyperbolic. Symmetric
hyperbolicity is yet a stronger condition than strong hyperbolicity.

The first, crucial theorem is that the initial value problem for the
system~\eqref{eqn:FT1S_CC} is well-posed in~$L^2$
\begin{align}
  ||\mathbf{u}(t,\cdot)||_{L^2}^2
  =\int_{\Sigma_t}\textrm{d}x\,\mathbf{u}(t,x)^T\mathbf{u}(t,x)
\end{align}
if and only if it is strongly hyperbolic. We denote the transpose by
superscript~$T$. Interestingly, if the system is only weakly
hyperbolic, it may be well-posed in a norm that contains specific
derivatives of the certain components of the state vector. In contrast
to~$L^2$ well-posedness for strongly hyperbolic systems however, the
weakly hyperbolic system is sensitive to changes in the source
terms~$\mathbf{s}$, and well-posedness is therefore delicate to verify
in practice. The second important result is that, for the initial
boundary value problem, which consists of specifying initial data on a
partial Cauchy slice at~$t=0$ together with boundary conditions at the
boundary of the domain at each time, there is a simple prescription to
choose boundary conditions such that the problem becomes well-posed
when the system~\eqref{eqn:FT1S_CC} is symmetric hyperbolic. This
straightforward approach is made possible by the fact that symmetric
hyperbolic systems admit energy estimates
like~\eqref{eqn:WP_CC_Energy_Estimate} by simple physical space
arguments involving the divergence theorem, without having to use the
Fourier-transform. We demonstrate this below for the wave equation. In
GR the application of this result for the initial boundary value
problem is greatly complicated by the constraints. The reason for this
is that, just as we need to choose initial data that satisfies the
constraints, so do we have to choose boundary conditions that are
constraint preserving. Likewise, boundary conditions need to be chosen
so that they model the physics we are interested in studying. For
instance, reflecting boundary conditions are less favored than those
which can absorb outgoing waves.

There are obvious differences between the simple system considered in
this section and the field equations of GR. First, the formulations of
GR we have seen are not first order in all derivatives but rather,
first order in time and second order in space. This is purely a
technical detail, however, which can be overcome introducing a first
order reduction and then applying the definitions as given here. An
elementary example of this is given already by the wave equation,
discussed momentarily, for which it is easy to see that there is a
symmetric hyperbolic first order reduction. Second, the Einstein
equations are not linear. This can be surmounted by linearizing about
an arbitrary background and then working in a small neighborhood of a
point, where to a good approximation the coefficients in the
linearized problem will be constant, and we can apply the definitions
given in this section. This is known as the localization principle.
With additional smoothness assumptions, strong hyperbolicity still
guarantees {\it local well-posedness} in a certain norm. The price we
pay for linearization and working just in the neighborhood of each
point is that, without investing much more work, we can only guarantee
existence of the solution for a short time. Naturally, long-term
behavior of solutions is a relevant question of great interest, and
one on which much progress is being made. But from a pure NR
perspective, it is the property of local well-posedness that
determines whether or not we can hope to accurately approximate
solutions computationally. There is much to say about all this, but to
avoid a lengthy discussion, we direct the interested reader to the
references highlighted above.

The canonical example of a PDE with a well-posed Cauchy problem is the
wave equation in flat-space, which we now consider as a simple
example. The wave equation is given by
\begin{align}
  \Box_\eta\phi&=\eta^{ab}\nabla_a\nabla_b\phi=-\p_t^2\phi+\p^i\p_i\phi=0\,,
  \label{eqn:WE}
\end{align}
where here~$\Box_\eta$ is the d'Alembert operator associated with the
Minkowski metric, $\nabla_a$ is the Levi-Civita derivative
of~$\eta_{ab}$ and in the second equality we assume that the
coordinates are global inertial. We can draw a crude analogy between
GR and the wave equation in which this is identified with the fully
second order form of the field equations~\eqref{eqn:EEs}. Introducing
the time reduction variable~$\pi=-\p_t\phi$, we can write the system
in first-order-in-time-second-order-in-space form
\begin{align}
  \p_t\phi&=-\pi\,,\quad \p_t\pi=-\p^i\p_i\phi\,.
  \label{eqn:FT2S_WE}
\end{align}
In analogy to our treatment of GR, this is like the evolution
equations~\eqref{eqn:ADM}. We can then reduce the system to first
order by introducing the spatial reduction variable~$\Phi_i$ and the
reduction constraint
\begin{align}
C_i=\p_i\phi-\Phi_i=0\,.\label{eqn:WE_reduction_constraint}
\end{align}
Even in the linear constant coefficient setting, not every PDE will
admit a first order reduction of the type~\eqref{eqn:FT1S_CC}. Such
PDEs are not hyperbolic. The canonical example of this is the heat
equation~$\p_t\phi=\p^i\p_i\phi$. In first order form, the wave
equation can be rewritten as
\begin{align}
  \p_t\phi=-\pi\,, \quad
  \p_t\Phi_i=-\p_i\pi\,, \quad
  \p_t\pi=-\p^i\Phi_i\,.\label{eqn:WE_FT1S}
\end{align}
Solutions of the reduction agree with those of the wave equation
whenever the reduction constraint~\eqref{eqn:WE_reduction_constraint}
is satisfied. We did not introduce a first order reduction of GR, but
can perfectly easily do so. In common with what we saw above for GR
however, if the reduction constraint is initially satisfied, it will
remain so, since we can easily show that~$\p_tC_i=0$. Taking the state
vector to be~$\mathbf{u}=(\phi,\Phi_i,\pi)^T$, in the notation
of~\eqref{eqn:FT1S_CC}, we have
\begin{align}
  \mathbf{A}^p=
  \begin{pmatrix}
    0 & 0 & 0\\
    0 & 0 & -\delta_i{}^p\\
    0 & -\delta^{jp} & 0
  \end{pmatrix}\,, \quad
  \mathbf{s} =
  \begin{pmatrix}
    -\pi \\0 \\ 0
  \end{pmatrix}\,.
\end{align}
The principal symbol~$\mathbf{P}^s$ is then
\begin{align}
  \mathbf{P}^s=\mathbf{A}^ps_p=
  \begin{pmatrix}
    0 & 0 & 0\\
    0 & 0 & -s_i\\
    0 & -s^j & 0
  \end{pmatrix}\,. 
\end{align}
It has real eigenvalues~$0,\pm1$, and so the reduction is at least
weakly hyperbolic. The eigenvectors of~$\mathbf{P}^s$ are
complete. They are given by
\begin{align}
  \mathbf{u}_0=(1,0_j,0)^T\,,
  \quad \mathbf{u}_A=(0, v^A_{j}, 0)\,,
  \quad \mathbf{u}_\pm= (0,s_j, \pm 1)\,,
\end{align}
where we have introduced two orthogonal unit covectors~$v^A_j$
for~$A=1,2$ that are orthogonal to~$s_i$. The uniformity
condition~\eqref{eqn:SH_uniformity_condition} is satisfied
with~$K=1$. In this case the principal matrices are already symmetric,
so we can choose the symmetrizer~$\mathbf{H}=\mathbf{1}$,
the~$5\times5$ identity.

Consider the following norm for our first order reduction of the wave
equation
\begin{align}
E(t,\mathbf{u})
=\int_{\Sigma_t}\textrm{d}x\,\mathbf{u}(t,x)^T\mathbf{H}\mathbf{u}(t,x)\,.
\label{eqn:FT1S_WE_Energy}
\end{align}
In the present context this corresponds almost exactly to the physical
energy of the solution. For this reason estimates that hold for such
norms are often referred to as energy estimates. Taking a time
derivative of~\eqref{eqn:FT1S_WE_Energy}, substituting the equations
of motion and rearranging, we obtain
\begin{align}
\p_tE(t,\mathbf{u})\leq E(t,\mathbf{u}) 
+ \int \textrm{d}x\,\p_p
\left(\mathbf{u}(t,x)^T\mathbf{H}\mathbf{A}^p\mathbf{u}(t,x)\right) \,.
\label{eqn:FT1S_WE_Esimate_dot}
\end{align}
Using the divergence theorem to remove the second term, and a simple
application of Gr\"onwall's lemma, we conclude with a well-posedness
estimate
\begin{align}
  E(t,\mathbf{u})^{1/2}\leq e^{t/2} E(0,\mathbf{f})^{1/2}\,,
\label{eqn:FT1S_WE_Esimate}
\end{align}
with initial data~$\mathbf{u}(0,x)=\mathbf{f}(x)$. This is precisely of the
form~\eqref{eqn:WP_CC_Energy_Estimate} in the~$L^2$ norm. The
computations to obtain estimates for an arbitrary symmetric hyperbolic
PDE are identical in spirit.

There are various PDE setups of interest in NR besides the initial and
initial boundary value problems. Most important are the characteristic
initial and characteristic initial boundary value problems, and the
hyperboloidal initial value problem. For an introduction to these,
see~\cite{Win05,Fra04,Val16}.

In the following section, besides presenting the equations themselves,
we give a summary of the status of the initial and initial boundary
value problems for the most popular formulations of GR in use in NR.

\subsection{Free-Evolution Formulations of GR}
\label{Sec:Free-Evolution}

We saw in section~\ref{Sec:3+1} that the constraint subsystem for the
ADM evolution equations~\eqref{eqn:ADM} is closed. If we modify the
evolution equations by adding combinations of the constraints, this
property still holds. We also saw that the field equations of GR do
not by themselves impose a choice for the lapse and shift. Thus there
is a large freedom in the choice of equations to be solved. A concrete
choice for these is known as a (free-evolution) formulation of GR. We
observed in section~\ref{Sec:WP} that for sufficiently smooth initial
data, strong hyperbolicity guarantees local well-posedness of the
Cauchy problem. Taking these considerations together, we are guided
when constructing formulations of GR to impose {\it at least} strong
hyperbolicity. Such a formulation is often called a hyperbolic
reduction. But there are issues besides this. It is highly desirable
to obtain gauge conditions that are flexible enough to reliably treat
a large range of scenarios of interest. A clever choice of evolved
variables can moreover help to reduce numerical error in applications,
or even to make it possible to treat data otherwise out of reach.

The basic ingredients in the construction of free-evolution
formulations are first, to make a choice of gauge, second, the
definition of auxiliary constraints, and finally the choice of
coupling between the field equations and the complete set of
constraints. These are already well-illustrated by the archetypal
example of a hyperbolic formulation of GR, the generalized harmonic
gauge formulation. To construct this, let us begin with the
trace-reversed Einstein equations written out, in coordinates~$x^\mu$
defined as in section~\ref{Sec:3+1}, in the form,
\begin{align}
  R_{\alpha\beta}
  &= -\tfrac{1}{2}g^{\mu\nu}\p_\mu\p_\nu g_{\alpha\beta}
    + \p_{(\alpha}{}^{\tiny{(4)}}\Gamma_{\beta)}
    - {}^{\tiny{(4)}}\Gamma^{\gamma}{}_{\alpha\beta}
    {}^{\tiny{(4)}}\Gamma_{\gamma}
    \nonumber\\
  &\quad
    + g^{\mu\nu} g^{\delta\epsilon} \left(\p_\delta g_{\mu\alpha}
    \p_\epsilon g_{\nu\beta}
    - {}^{\tiny{(4)}}\Gamma_{\alpha\mu\delta}
    {}^{\tiny{(4)}}\Gamma_{\beta\nu\epsilon}\right)
    \nonumber\\
  &= 8\pi\big(T_{\alpha\beta} - \tfrac{1}{2}g_{\alpha\beta}T\big) \,,
    \label{eqn:Trace-Reversed-EEs_expanded}
\end{align}
where we introduce the contracted Christoffel
symbols~${}^{\tiny{(4)}}\Gamma_{\alpha}
=g^{\mu\nu}\,{}^{\tiny{(4)}}\Gamma_{\alpha\mu\nu}$. Comparing the
principal part with the wave equation~\eqref{eqn:WE}, we see that
there would be a strong structural similarity if only the second term
in the right-hand side, the gradient of the contracted Christoffel
symbol were absent. Of course we are not free to simply change the
equations to be solved. But looking more closely we realize that the
contracted Christoffel symbol can be written as,
\begin{align}
  {}^{\tiny{(4)}}\Gamma_{\alpha}&=- g_{\alpha\beta} \Box_g x^\beta\,,
\end{align}
with~$\Box_g=\nabla^a\nabla_a$ here the curved space
d'Alembertian. Therefore, we might work with generalized harmonic,
also called generalized wave, coordinates
\begin{align}
  \Box_g x^\alpha=H^\alpha(x,g)\,.
\end{align}
When the gauge source functions~$H^\alpha(x,g)$ vanish, these are
called harmonic coordinates or wave coordinates. Substituting this
expression back into~\eqref{eqn:Trace-Reversed-EEs_expanded} directly
and using the fact that the gauge source functions depend upon the
metric, but not partial derivatives of the metric components, the
principal part of the equations indeed agrees with that of the curved
space wave equation. Formally this is achieved by defining the
auxiliary Harmonic constraints
\begin{align}
  C_\alpha = {}^{\tiny{(4)}}\Gamma_{\alpha}+H_{\alpha} = 0\,,
\end{align}
and replacing the Einstein equations by the reduced field equations
\begin{align}
  \mathcal{R}_{\alpha\beta}&=R_{\alpha\beta}-\nabla_{(\alpha} C_{\beta)}
                             \nonumber\\
  &= -\tfrac{1}{2}g^{\mu\nu}\p_\mu\p_\nu g_{\alpha\beta}
    + \nabla_{(\alpha}H_{\beta)}\nonumber\\
  &\quad
    + g^{\mu\nu} g^{\delta\epsilon} \left(\p_\delta g_{\mu\alpha}
    \p_\epsilon g_{\nu\beta}
    - {}^{\tiny{(4)}}\Gamma_{\alpha\mu\delta}
    {}^{\tiny{(4)}}\Gamma_{\beta\nu\epsilon}\right)
    \nonumber\\
  &= 8\pi\big(T_{\alpha\beta} - \tfrac{1}{2}g_{\alpha\beta}T\big) \,,
    \label{eqn:Trace-reversed-Reduced_EEs}
\end{align}
where~$\mathcal{R}_{\alpha\beta}$ is called the reduced Ricci
tensor. This is called the generalized harmonic gauge (GHG)
formulation of GR. Due to the similarity with the wave equation
discussed in the last section, it is not surprising that the GHG
formulation admits a symmetric hyperbolic first order reduction, and
thus has a well-posed Cauchy problem. The terminology `reduced' arose
as an abbreviation of hyperbolic reduction. To see that the constraint
subsystem remains closed, we reverse the trace once more to obtain
\begin{align}
  G_{ab}-\nabla_{(a}C_{b)}+\tfrac{1}{2}g_{ab}(\nabla_cC^c)
  =8\pi T_{ab}\,.\label{eqn:Reduced_EEs}
\end{align}
Contracting then with the normal vector~$n^a$ and employing the
language of the~$3+1$ decomposition once more, a short calculation
gives
\begin{align}
  \Lie_tC_n &= \alpha \big[H-D_a(\!\perp\!\!C^a)
  +a^a(\!\perp\!\!C_a)-2KC_n\big]
  +\Lie_\beta C_n\,,\nonumber\\
  \perp\!\Lie_t(\!\perp\!\! C)_a &= \alpha[2M_a-D_aC_n
  +a_aC_n-2K_a{}^b(\!\perp\!\! C)_b]
  +\perp\!\Lie_\beta(\!\perp\!\! C)_a\,,
  \label{eqn:Harmonic_Constraint_Dot}
\end{align}
for~$C_n=C_an^a$ and~$\!\perp\!\!C_a$, so we see that time derivatives
of the harmonic constraints are essentially given by the Hamiltonian
and momentum constraints we met in section~\ref{Sec:ID}. Taking the
divergence of~\eqref{eqn:Reduced_EEs} and using local stress-energy
conservation we have
\begin{align}
 \Box C_a + R_{ab}C^b = 0 \,,
\end{align}
where we can now substitute back
from~\eqref{eqn:Trace-reversed-Reduced_EEs} to remove the Ricci
tensor. This means that the harmonic constraints satisfy a second
order nonlinear wave equation in which every term appears with at
least one constraint. Combining this
with~\eqref{eqn:Harmonic_Constraint_Dot} we can conclude that if we
start with constraint satisfying initial data, the time development,
which is guaranteed to exist, at least locally, by well-posedness,
will satisfy the constraints too.

We have presented GHG formulation itself by returning to the spacetime
view, but this was simply because the structure in the equations is
more evident in that notation. We can equivalently work with the~$3+1$
language. The harmonic constraints, for instance, can be rewritten as
\begin{align}
\p_t\alpha&=-\alpha^2(K + C_n + H_n)+\Lie_\beta\alpha\,,\nonumber\\
  \p_t\beta^i&=\alpha^2(\Gamma^i+ (\!\perp\!\!C)^i+(\!\perp\!\!H)^i)
               -\alpha D^i\alpha
             +\beta^j\p_j\beta^i\,.
\end{align}
So it turns out that a choice of gauge as a second order wave equation
on the coordinates corresponds to a choice of first order equations of
motion for the lapse and shift in the~$3+1$ setting, consistent with
the interpretation we gave earlier that the lapse and shift encode the
freedom in choosing coordinates.

To build the reduced Einstein
equations~\eqref{eqn:Trace-reversed-Reduced_EEs}, we chose a specific
combination of derivatives of constraints to add to the Einstein
equations. As we have seen, local well-posedness of hyperbolic PDEs
only depends upon the principal part. Therefore one may wonder why we
chose to add a symmetrized covariant derivative, which contains
non-principal constraint additions. The answer in this case is that
the choice we made presently results in the simplest form of the
equations, which we hope makes the general strategy more
transparent. More generally, one might feel that the use of the
Levi-Civita derivative is preferred as it is covariant, but if we take
seriously the point of view of~\cite{Wal84} then the same applies to
partial derivatives. Nevertheless, it is perfectly legitimate to ask
whether one non-principal constraint addition or another is to be
preferred. In numerical applications, error will generically violate
the constraints. Although such violations should converge to zero with
increasing resolution, any strategy to suppress them at finite
resolution is welcome. The behavior of numerical error reflects to
some extent the behavior of the continuum equations. Motivated by this
type of reasoning, a crucial adjustment to the
formulation~\eqref{eqn:Trace-reversed-Reduced_EEs} for numerical
applications was obtained. The idea is to carefully make non-principal
constraint additions,
\begin{align}
  \mathcal{R}_{\alpha\beta}=R_{\alpha\beta}-\nabla_{(\alpha} C_{\beta)}
  + \gamma_0 \big(n_{(a}C_{b)}-\tfrac{1}{2}g_{ab}C_cn^c\big)
  = 8\pi\big(T_{\alpha\beta} - \tfrac{1}{2}g_{\alpha\beta}T\big) \,.
  \label{eqn:Trace-reversed-Reduced_damping_EEs}
\end{align}
Linearizing, it can be seen that choosing~$\gamma_0$ positive leads to
the exponential suppression of small, high-frequency constraint
violations. Combined with the wavelike propagation of the harmonic
constraints, this idea is essentially the same as that of hyperbolic
divergence cleaning in GRMHD. The use of such constraint damping
schemes is now standard with various formulations of GR. Besides this
constraint damping scheme, there are various different choices of
non-principal constraint addition in use in the NR literature, giving
rise to slightly different flavors of the GHG formulation.

The GHG formulation as presented here is often subjected to a first
order in time (or even fully first order) reduction when implemented
in practice. Together these different flavors of GHG constitute one of
the two main types of formulations of GR in regular use for
astrophysical work in NR. The GHG formulation is symmetric
hyperbolic. As discussed above, this allows for much more
straightforward PDE analysis. In particular, this has led to a proof
of well-posedness of the initial boundary value problem with boundary
conditions that are constraint preserving and gravitational radiation
controlling. The GHG formulation is the only formulation of GR with
this property in frequent use in NR simulations that actually uses
these boundary conditions in applications.

We now come to the second class of formulations of GR in frequent use
for astrophysical NR. In the literature these are referred to as the
Baumgarte-Shapiro-Shibata-Nakamura-Oohara-Kojima (BSSNOK or BSSN),
CCZ4 and Z4c formulations. All three are constructed following the
general template laid out above. They are furthermore closely related
and indeed, the latter two are essentially variations of each other,
as they share principal part and constraint damping terms. In
particular, all three formulations share in common the use of a
conformal decomposition in the choice of variables. We will try to
give a unified presentation that covers all three simultaneously. We
begin with the trace-reversed Einstein equations and, as above, define
auxiliary constraints, this time called~$Z_a$. We then replace the
Einstein equations by the Z4 equations of motion
\begin{align}
  R_{ab}+2\nabla_{(a}Z_{b)}
  -\kappa_1 \big[ 2 n_{(a} Z_{b)} - (1+\kappa_2)g_{ab}n_c Z^c \big]
  +W_{ab}(Z)
  &=8\pi\big(T_{ab}-\tfrac{1}{2}g_{ab}T\big)\,,\label{eqn:4d_Z4}
\end{align}
where~$W_{ab}(Z)$ is a symmetric tensor that vanishes when~$Z_a=0$,
but which is otherwise arbitrary. This tensor parametrizes the
different flavors of the formulation. As in the GHG formulation, we
have introduced constraint damping terms with parameters~$\kappa_1$
and~$\kappa_2$. Solutions of these equations agree with those of GR
when~$Z_a=0$.

Comparing the Z4 equations of motion
with~\eqref{eqn:Trace-reversed-Reduced_damping_EEs}, there is an
obvious similarity. In fact, choosing
\begin{align}
  Z_a=-\tfrac{1}{2}C_a\,,\quad \kappa_1 = \gamma_0\,, \quad \kappa_2=0\,,
\end{align}
and taking~$W_{ab}=0$, the expressions align perfectly. The important
difference, however, is that in the Z4 setup we have not yet chosen an
equation of motion for the lapse and shift. Morally, therefore, we may
think of Z4 as a formulation with the same constraint subsystem as
GHG, but without imposing generalized harmonic coordinates.

We will follow the common presentation of these conformal variable
formulations in using the~$3+1$ language and working in adapted
coordinates, writing the components of the equations of motion. We
begin by decomposing the Z4 constraint, introducing~$\Theta=-n_aZ^a$,
and the spatial part~$Z_i$.  Geometrically this is defined by
application of the projection operator, but in adapted coordinates we
can use the components downstairs. The definition of the extrinsic
curvature remains unchanged, so we have
\begin{align}
  \p_t\gamma_{ij}&=-2\alpha K_{ij}+\Lie_\beta\gamma_{ij}\,,
  \label{eqn:ADM_Z4_gamma}
\end{align}
as in the ADM setup above. For the extrinsic curvature itself we
simply take the spatial part of~\eqref{eqn:4d_Z4}. This of course
gives equations that look like the ADM evolution equations
for~$K_{ij}$, plus constraint additions,
\begin{align}
  \p_tK_{ij}&=-D_iD_j\alpha
              +\alpha\left[R_{ij}-2K^k{}_iK_{jk}+KK_{ij}+2D_{(a}Z_{b)}
              -2K_{ij}\Theta+W_{ij}\right]\nonumber\\
            &\quad -\alpha\kappa_1(1+\kappa_2)\gamma_{ij}\Theta
              + 4 \pi \alpha\left[\gamma_{ij}(s-\rho)-2s_{ij}\right]
              +\Lie_\beta K_{ij} \,.
              \label{eqn:ADM_Z4_K}
\end{align}
For the constraints, we trace-reverse~\eqref{eqn:4d_Z4} and contract
with~$n^a$ twice (for the~$\Theta$ equation of motion) and likewise
contract once with~$n^a$ and once with the projection operator (for
the~$Z_i$ equation). This gives,
\begin{align}
  \p_t\Theta &= \alpha\big[ \tfrac{1}{2} H +D^iZ_i
               -a^iZ_i - K\Theta -\kappa_1(2+\kappa_2)\Theta
               + \hat{W}_{nn} \big]
               +\Lie_\beta\Theta \,,\nonumber\\
  \p_tZ_i&= \alpha \big[ M_i + D_i\Theta -
           a_i\Theta -2K_i{}^jZ_j -\kappa_1 Z_i -W_{ni} \big] 
           +\Lie_\beta Z_i \,. \label{eqn:ADM_Z4_Theta_Z}
\end{align}
Where we denote~$\hat{W}_{nn}=W_{nn}-W/2$. The complete collection of
the constraints is then given by
\begin{align}
  \Theta=0\,, \quad Z_i=0\,, \quad H=0\,, \quad  M_i=0\,,
  \label{eqn:Z4_Constraints}
\end{align}
with the Hamiltonian and momentum constraints defined exactly as
earlier in~\eqref{eqn:Ham} and~\eqref{eqn:Mom}.

To derive the formal evolution equations for the Hamiltonian and
momentum constraints, we trace-reverse~\eqref{eqn:4d_Z4} and take the
divergence with~$\nabla_a$, then~$3+1$ decompose. Although
straightforward, this is considerably lengthier than for the ADM
equations. This reveals
\begin{align}
  \p_tH &= -2 \alpha D^iM_i-4M_iD^i\alpha+2\alpha K H
          +4\alpha(K\gamma^{ij}-K^{ij})
          (D_iZ_j-K_{ij}\Theta+\tfrac{1}{2}W_{ij})\nonumber\\
        &\quad-4\alpha\kappa_1(1+\kappa_2)K\Theta+\Lie_\beta H\,,
          \nonumber\\
  \p_tM_i&= -\tfrac{1}{2}\alpha D_iH+\alpha K M_i- (D_i\alpha) H
           + D^j\big(\alpha\big(2D_{(i}Z_{j)}-2K_{ij}\Theta
           +W_{ij}\big)\big)
           \nonumber\\
        &\quad -D_i\big(\alpha\big(2D_jZ^j-2K\Theta
          +\gamma^{jk}W_{jk}\big)\big)
          +2D_i\big(\alpha\kappa_1(1+\kappa_2)\Theta\big)
          +\Lie_\beta M_i\,. \label{eqn:ADM_Z4_H_M}
\end{align}

As discussed above, the actual evolved variables involve a conformal
decomposition not dissimilar to, and historically inspired by, that
employed in the treatment of the constraints as outlined in
section~\ref{Sec:ID}. The variables themselves are defined through
\begin{align}
  \chi &= \gamma^{-1/3}\,, 
 \quad 
  &\tilde{\gamma}_{ij}&=\gamma^{-1/3}\gamma_{ij}\,,  
  \quad
  &\hat{K} &= \gamma^{ij}K_{ij}-2\Theta\,,\nonumber\\
  \tilde{A}_{ij} &= \gamma^{-1/3}(K_{ij}-\tfrac{1}{3}\gamma_{ij}K) \,, 
  \quad
  &\tilde{\Gamma}^i&= 2\tilde{\gamma}^{ij}Z_j 
   +\tilde{\Gamma}_{\textrm{d}}{}^i  \,, 
  \quad
  &\tilde{\Gamma}_{\textrm{d}}{}^i &= 
  \tilde{\gamma}^{jk}\tilde{\Gamma}^i{}_{jk}
  =\tilde{\gamma}^{ij}\tilde{\gamma}^{kl}\p_k\tilde{\gamma}_{lj} \,.
\label{eqn:conformal_variables}
\end{align} 
There are minor changes in the choice of variables from one
formulation to another, and indeed across different implementations.
The key benefit of this change of variables is that they help in the
treatment of data in which the metric components blow-up in an
isotropic manner. Consider for instance the prototype for
puncture-type initial data, the Schwarzschild metric with the usual
time coordinate and isotropic coordinates. In that case, all
non-vanishing component of the spatial metric~\eqref{eqn:SS_isotropic}
diverge as~$r\to0$, but the evolved metric remains trivial. One would
therefore expect that perturbations would give a conformal metric that
could also be inverted without problem. This turns out to be the case
in a large variety of situations of interest. Nevertheless, since the
determinant of the metric diverges we must still manage one singular
variable. In practice this is managed by evolving~$\chi$, a negative
power of the determinant of the spatial metric, see the definition
in~\eqref{eqn:conformal_variables}. The definition of~$\chi$ thus
guarantees that the evolved variable remains finite, although keeping
in mind~\eqref{eqn:SS_isotropic} as a model, we may be concerned about
the degree of smoothness of solutions when~$\chi$ vanishes, and
moreover any inverse power of~$\chi$ that appears in the equations of
motion. The remaining variables are chosen so that the complete set of
equations of motion appear as regular as possible. (There is no proof
that this is the optimal choice, but many different choices have been
investigated. With carefully chosen numerical methods, even less
regular choices of variable may be managed, but the conformal
decomposition remains the most prevalent choice to manage this type of
data). We see that under this change we have apparently increased the
number of evolved variables, but we have also introduced the algebraic
constraints
\begin{align}
  \det\tilde{\gamma}_{ij}=\tilde{\gamma}=1\,, \qquad
  \tilde{\gamma}^{ij}\tilde{A}_{ij}=\tilde{A}=0\,.
  \label{eqn:Conformal_Z4_algebraic}
\end{align}
When these constraints are explicitly enforced, which is typically
done in practice, the solution spaces of the evolution equations with
either the ADM or the conformal variables are isomorphic and they
therefore have shared PDE properties.

Pushing the evolution equations~\eqref{eqn:ADM_Z4_gamma}, through the
change of variables, we obtain
\begin{align}
  \p_t\chi &= \frac{2}{3}\chi \big[ \alpha(\hat{K}+2\Theta)-D_i\beta^i \big]
             \,,\nonumber\\
  \p_t\tilde{\gamma}_{ij} &=-2\alpha \tilde{A}_{ij}+\beta^k\p_k\tilde{\gamma}_{ij}+
                            2\tilde{\gamma}_{k(i}\p_{j)}\beta^k 
                            -\frac{2}{3} \tilde{\gamma}_{ij}\p_k\beta^k \,,
                            \label{eqn:Conformal_Z4_gamma}
\end{align}
Likewise from~\eqref{eqn:ADM_Z4_Theta_Z} we have
\begin{align}
  \p_t\Theta &= \alpha\big[ \tfrac{1}{2} H +D^iZ_i
               -a^iZ_i - K\Theta -\kappa_1(2+\kappa_2)\Theta
               + \hat{W}_{nn} \big]
               +\Lie_\beta\Theta \,,\label{eqn:Conformal_Z4_Theta}
\end{align}
for the~$\Theta$ constraint and the complicated expression
  \begin{align}
  \p_t\tilde{\Gamma}^i&= 2\alpha\big[
                        \tilde{\Gamma}^i{}_{jk}\tilde{A}^{jk} 
                        - \tfrac{3}{2}\tilde{A}^{ij}\p_j\ln\chi
                        - \tilde{A}^{ij}a_j-4\pi\tilde{\gamma}^{ij}s_j
                        -\tfrac{1}{3}\tilde{\gamma}^{ij}\p_j(2\hat{K}+\Theta)
                        \nonumber\\
                      &\quad -\tfrac{1}{3}(\hat{K}+2\Theta)(\tilde{\Gamma}^i
                        -\tilde{\Gamma}_{\textrm{d}}{}^i)
               -\tilde{\gamma}^{ij}a_j\Theta - \tilde{\gamma}^{ij}W_{jn}\big]
               -\alpha\kappa_1(\tilde{\Gamma}^i-\tilde{\Gamma}_{\textrm{d}}{}^i)
               \nonumber\\
             &\quad+\tilde{\gamma}^{jk}\p_j\p_k\beta^i
               +\tfrac{1}{3}\tilde{\gamma}^{ij}\p_j\p_k\beta^k
               +\beta^j\p_j\tilde{\Gamma}^i
               -\tilde{\Gamma}^j\p_j\beta^i
               +\tfrac{2}{3}\tilde{\Gamma}^i\p_j\beta^j\,,
             \label{eqn:Conformal_Z4_Z}
\end{align}
when we combine derivatives of~\eqref{eqn:Conformal_Z4_gamma}
with~\eqref{eqn:ADM_Z4_Theta_Z}. Finally, doing the same
for~\eqref{eqn:ADM_Z4_K}, we arrive at
\begin{align}
  \p_t\hat{K} &= -D^iD_i\alpha
                +\alpha[\tilde{A}^{ij}\tilde{A}_{ij}
                +\tfrac{1}{3}(\hat{K}+2\Theta)^2]  
                +4\pi\alpha [s+\rho]\nonumber\\
              &\quad
                +\alpha[2a^iZ_i+K\Theta
                +2\kappa_1(1+\kappa_2)\Theta-W_{nn}]
                +\Lie_\beta\hat{K}\,,\label{eqn:Conformal_Z4_K}
\end{align}
for the trace variable, and
\begin{align}
  \p_t\tilde{A}_{ij} &= \chi\big[-D_iD_j\alpha
                       +\alpha(R_{ij}+2D_{(i}Z_{j)}
                       +W_{ij}-8\pi s_{ij})\big]^{\textrm{tf}}
                       +\alpha\big[\hat{K}\tilde{A}_{ij}
                       -2\tilde{A}^k{}_i\tilde{A}_{jk}\big]\nonumber\\
                     &\quad + \beta^k\p_k\tilde{A}_{ij}
                       + 2\tilde{A}_{k(i}\p_{j)}\beta^k 
                       - \tfrac{2}{3}\tilde{A}_{ij}\p_k\beta^k\,,
                       \label{eqn:Conformal_Z4_A}
\end{align}
for the trace-free part of the extrinsic curvature. Here we have
employed a number of auxiliary definitions and shorthands. The
superscript~$\textrm{tf}$ denotes the trace-free part as usual. The
spatial Christoffel symbols can be expressed in terms of the conformal
connection via,
\begin{align}
  \Gamma^k{}_{ij}&=\tilde{\Gamma}^k{}_{ij}-\delta^k{}_{(i}\p_{j)}\ln\chi
                   +\tfrac{1}{2}\tilde{\gamma}_{ij}
                   \tilde{\gamma}^{kl}\tilde{\p}_l\ln\chi\,.
\end{align}
In terms of the conformal variables, the spatial Ricci tensor can be
expressed as
\begin{align}
  R_{ij}&=\tilde{R}_{ij}^\chi+\tilde{R}_{ij}\,,\nonumber\\
  \tilde{R}_{ij}^\chi&= \tfrac{1}{2\chi}\tilde{D}_i\tilde{D}_j\chi
                       +\tfrac{1}{2\chi}\tilde{\gamma}_{ij}
                       \tilde{D}^k\tilde{D}_k\chi
                       -\tfrac{1}{4\chi^2}\tilde{D}_i\chi\tilde{D}_j\chi
                       -\tfrac{3}{4\chi^2}\tilde{\gamma}_{ij}
                       \tilde{D}^k\chi\tilde{D}_k\chi
                       \,,\nonumber\\
  \tilde{R}_{ij}&= -\tfrac{1}{2}\tilde{\gamma}^{kl}\p_k\p_l\tilde{\gamma}_{ij} 
                  +\tilde{\gamma}_{k(i}\p_{j)}\tilde{\Gamma}_{\textrm{d}}{}^k 
                  +\tilde{\Gamma}_{\textrm{d}}{}^k\tilde{\Gamma}_{(ij)k}
                  +\tilde{\gamma}^{lm}\left(2\tilde{\Gamma}^k{}_{l(i}
                  \tilde{\Gamma}_{j)km}
                  +\tilde{\Gamma}^k{}_{im}\tilde{\Gamma}_{klj}\right) \,,
                  \label{eqn:Z4_Spatial_Curvature}  
\end{align}

To summarize, in terms of the conformal variables the Z4 constraints
are
\begin{align}
  \Theta=0\,,\qquad
  2Z_i=\tilde{\gamma}_{ij}(\tilde{\Gamma}^j-\tilde{\Gamma}_\textrm{d}{}^j)\,.
  \label{eqn:Conformal_Z4_Constraints_Theta_Z}
\end{align}
The Hamiltonian and momentum constraints are
\begin{align}
  H&=R-\tilde{A}^{ij}\tilde{A}_{ij}+\tfrac{2}{3}(\hat{K}+2\Theta)^2
     -16\pi\rho=0\,,\nonumber\\
  \tilde{M}^i&=\tilde{\gamma}^{ij}M_j=\tilde{D}_j\tilde{A}^{ij}
                          -\tfrac{2}{3}\tilde{D}^i(\hat{K}+2\Theta)
                          -\tfrac{3}{2}\tilde{A}^{ij}\tilde{D}_j\ln\chi
                          -4\pi\tilde{\gamma}^{ij}s_j=0\,,
                          \label{eqn:Conformal_Z4_Constraints_Ham_Mom}
\end{align}
in addition to which we have the algebraic
constraints~\eqref{eqn:Conformal_Z4_algebraic}, which are assumed to
be imposed for the reasons discussed above.

The most popular gauge condition in use with these conformal
formulations can be written
\begin{align}
  \p_t\alpha&= -2\alpha\hat{K}+\Lie_\beta\alpha\,,\nonumber\\
  \p_t\beta^i&=\mu_S\tilde{\Gamma}^i-\eta \beta^i + \beta^j\p_j\beta^i\,.
               \label{eqn:puncture_gauge}
\end{align}
This is known as the moving-puncture gauge. Again, there are various
different flavors in use, and here we have selected just the
simplest. Observe that just as in the GHG formulation, we choose here
evolution equations for the lapse and shift, and in fact these look
structurally similar to the generalized harmonic choice. The lapse
condition is referred to as the `$1+\log$' variant of the Bona-Mass\'o
slicing condition. The shift condition is known as `Gamma-driver
shift'. In GHG, all of the speeds of propagation associated with the
gauge choice coincide with the speed of light. That is not the case in
moving puncture gauge, which instead has various superluminal speeds.

Moving now to the constraint addition tensor~$W_{ab}$, observe that we
have
\begin{align}
  W_{ab}&= n_an_b W_{nn}-2n_{(a}\!\perp\!W_{b)n}
          +\tfrac{1}{3}\gamma_{ab}[2\hat{W}_{nn}-W_{nn}]
          +\!\perp\!W_{ab}^{\textrm{tf}}\,.
\end{align}
The first of the three common flavors of this `conformal
decomposition' setup, which we give in reverse historical order of
their appearance, is called CCZ4. In CCZ4 the only non-trivial piece
of the constraint addition tensor~$W_{ab}$ is taken to be
\begin{align}
  W_{in}&= \alpha^{-1}(\kappa_3-1)\big(\tfrac{2}{3}Z_{i}\p_k\beta^k
          -\gamma_{ij}\gamma^{kl}Z_k\p_l\beta^j\big)
          \,,\label{eqn:CCZ4}
\end{align}
often with~$\kappa_3=1$, but occasionally with~$\kappa_3=1/2$. One
needs to check the literature for details in different cases. CCZ4 is
often used with~$\chi^{1/2}$ in place of~$\chi$ as an evolved
variable. The second variation is called Z4c, in which the constraint
addition tensor is
\begin{align}
  W_{nn}&=2a^iZ_i+K\Theta+\kappa_1(1+\kappa_2)\Theta\,,\nonumber\\
  \hat{W}_{nn}&=a^iZ_i+K\Theta+\Gamma^iZ_i
  -\tfrac{2}{3}\Gamma^i{}_{ij}Z^j\,,\nonumber\\
  \tilde{\gamma}^{ij}W_{jn}&=\tfrac{1}{3}(\hat{K}+2\Theta)
  (\tilde{\Gamma}^i-\tilde{\Gamma}_{\textrm{d}}{}^i)
  +\tilde{\gamma}^{ij}a_j
  -\alpha^{-1}\tilde{\gamma}^{jk}Z_k\p_j\beta^i
  +\tfrac{2}{3}\alpha^{-1}\tilde{\gamma}^{ij}Z_j\p_k\beta^k\,,\nonumber\\
  W_{ij}^{\textrm{tf}}&=2\Theta K_{ij}^{\textrm{tf}}
  +2\big[\Gamma^k{}_{ij}Z_k
  -\tfrac{2}{3}\Gamma^k{}_{k(i}Z_{j)}\big]^{\textrm{tf}}\,.
  \label{eqn:Z4c}
\end{align}
The reason for these choices was to make the equations of motion as
close as possible to those of the following formulation. To obtain the
the third popular option, the BSSN formulation (called on occasion
BSSNOK), one need only take the Z4c equations of motion, and set
\begin{align}
  \Theta=0\,,\label{eqn:Z4_to_BSSNOK}
\end{align}
discarding it as an evolved variable, and set all of the constraint
damping parameters~$\kappa_1,\kappa_2$ to zero. Further variations,
including first order reductions and slightly different choices of
variable have been investigated. Observe that here we have sacrificed
fidelity to the notation used in the literature to present all three
formulations together simultaneously.

With suitable choices for the gauge parameter~$\mu_S$
in~\eqref{eqn:puncture_gauge} all three variants are strongly
hyperbolic and therefore as discussed in section~\ref{Sec:WP} admit a
well-posed initial value problem. For the moving puncture gauge none
are symmetric hyperbolic, and so well-posedness results for the
initial boundary value problem are technically harder to come by. Such
results are well summarized in~\cite{SarTig12} (see
also~\cite{HilRui16} for a proof of well-posedness for a system close
to those presented here). Consequently most implementations working
with puncture gauge employ a naive outer boundary treatment, but try
to place the boundary sufficiently far away from the region of
interest that it has a negligible effect on the physics of the system.

In the principal part the relationship between gauge, constraints and
the gravitational wave degrees of freedom can be understood
straightforwardly. It turns out that hyperbolicity of gauge choices
can be understood independently of the formulation of GR, and that
strong hyperbolicity of a gauge choice is a necessary condition that a
formulation built with that gauge will itself be strongly
hyperbolic. The same holds for the constraint subsystem.

Supposing we want to tackle a particular physical problem, what
principles can we use to choose between different formulations, both
here and more generally? In accordance with our discussion above, a
first requirement is that of well-posedness of the initial boundary
value problem, ideally with boundary conditions that can be
implemented in practice. From this point of view the GHG formulation
is preferred over those which use moving puncture gauge. On the other
hand, as discussed momentarily, the use of the conformally decomposed
variables does give a clear advantage in managing the puncture
representation of black holes. As we have seen, there is little
difference between the different flavors of conformal decomposition of
Z4. Both have a constraint subsystem that looks like the wave equation
in the principal part, and accept the same constraint damping scheme
as the GHG formulation. Non-principal differences in the constraint
subsystem do make a difference to the long-term behavior of constraint
violations and can induce exponential growth or worse. The analysis
required to understand this for a particular constraint addition is,
unfortunately, technically involved, and so there is no systematic
understanding of what choices are optimal in general. The wavelike
nature and damping scheme for the Z4 constraint subsystem has been
shown to give advantages over BSSN computations in certain
scenarios. The reason for this is that the BSSN constraint subsystem
contains slow speeds, which result in constraint violation that sits
stationary on the numerical domain and grows.

A crucial feature of formulations of NR is in how black holes are to
be managed. We saw in our discussion of initial data that binary black
hole initial data can be modeled on the Schwarzschild solution in
isotropic coordinates. These are called puncture initial data. The
crucial strength of working with the conformally decomposed
variables~\eqref{eqn:conformal_variables} and the moving-puncture
gauge~\eqref{eqn:puncture_gauge} is that the points at which the
conformal factor~$\chi$ vanishes (the puncture) is naturally advected
around by the shift condition. In practice this results in a highly
versatile treatment with many different physical configurations that
requires little fine-tuning of the numerical algorithm. The
disadvantage of the moving-puncture approach is the limited regularity
of solutions at the puncture itself. Excision, in which the black hole
region is cut out of the computational domain explicitly, results in
smooth solutions but poses a much greater technical obstacle, since we
have to monitor the solution, either actively or passively, to be sure
that the excision boundary does not require boundary conditions and
that (some notion of) the black hole horizon remains within the
computational domain.

\section{The Physical Interpretation of Numerical Spacetimes}
\label{Sec:Interpretation}

In most cases the raw solution to the Einstein equations provided by
the metric over an extended region of spacetime does not correspond
directly to the physical quantity of interest. Instead the data is
subjected to post-processing. The key examples of this are black hole
horizons and gravitational waves. For brevity, we give only an
informal overview of each.

Asymptotically flat spacetimes can be characterized by the existence
of a certain region in which the metric can be represented as a
perturbation of the Minkowski metric in global inertial
coordinates. In these spacetimes there is a natural notion of future
null-infinity, the final destination for outgoing gravitational waves
emitted from the strong-field region. The black hole region, if it is
non-empty, is defined as the complement of the causal past of future
null-infinity. The boundary of the black hole region is called the
event horizon. This definition relies on global structure and is
therefore not applicable if we are given only a subset of the
spacetime without future null-infinity. Provided a large enough region
of data, we can nevertheless compute an accurate approximation to the
position of the event horizon. In practice this is done by integrating
null-geodesics backwards in time to find the points in spacetime
around which they accumulate. Since we compute spacetimes in NR
forward in time, this procedure has to be done by post-processing
data.

To understand the physics during the evolution itself it is often
helpful to have a proxy for the event horizon. For this we build on
the concept of a trapped surface. A trapped surface within a foliation
is a closed surface at a fixed instant of time~$\Sigma_t$ whose area
form would decrease from any point on the surface if we dragged that
point infinitesimally to the future along an outgoing null-geodesic
starting at said point. In case the area form would remain constant
under such a deformation, the surface is called marginally
trapped. The outermost marginally trapped surface is called the
apparent horizon. Trapped surfaces play an important role in the
singularity theorems. For the purposes of NR, the most relevant fact
is that if we have a spacetime that does not contain naked
singularities, as is expected generically in black hole spacetimes,
then any trapped surface must lie inside a cross-section of the event
horizon. The definition of the apparent horizon refers to instants of
time and so appears in different positions in different foliations,
but since it is defined locally in time can be used to characterize
black holes during a numerical evolution itself. For details of
numerical methods for apparent and event horizon searches,
see~\cite{Tho06}. For details of additional quasilocal quantities that
may be used to interpret spacetimes, see~\cite{Sza09}.

The primary deliverable of NR for gravitational wave astronomy is the
gravitational waveform emitted from a compact binary system. The
natural mathematical idealization of a distant observer in an
asymptotically flat spacetime, which serves as a model observer for
experiments, lies at future null-infinity. From the waveform one can
compute the total energy and linear momentum emitted from a system. As
is the case with the event horizon, since the computational domain
does not, in most cases, include future null-infinity, we can not
extract this signal from numerical data. Nevertheless, if a
sufficiently large domain is chosen an accurate approximation to the
signal at infinity can be computed by extrapolation. This wave
extraction procedure can be performed in multiple ways but the most
popular approach employs the Newman-Penrose formalism. Besides
textbook treatments, the review article~\cite{BisRez16} contains a
complete overview. There is an ongoing effort to include future
null-infinity in the computational
domain~\cite{Win05,HRN_web,ReiBisPol09,MaMoxSch23} in generic
asymptotically flat systems. Even with such an improvement,
calculating and extracting gravitational waveforms with NR will remain
too expensive, and too slow, to make an extremely dense sampling of
the solution space. For applications to gravitational wave astronomy,
accurate models that aggregate Post-Newtonian, perturbative and NR
information are constructed. For textbook introductions to
gravitational wave science and modeling,
see~\cite{Mag07a,CreAnd11,Mag18,And20}.

\section{Numerical Methods, Implementation and Codes}
\label{Sec:Numerical_Methods}

\begin{figure}[t!]
  \begin{center}
    \includegraphics[width=0.65\textwidth]{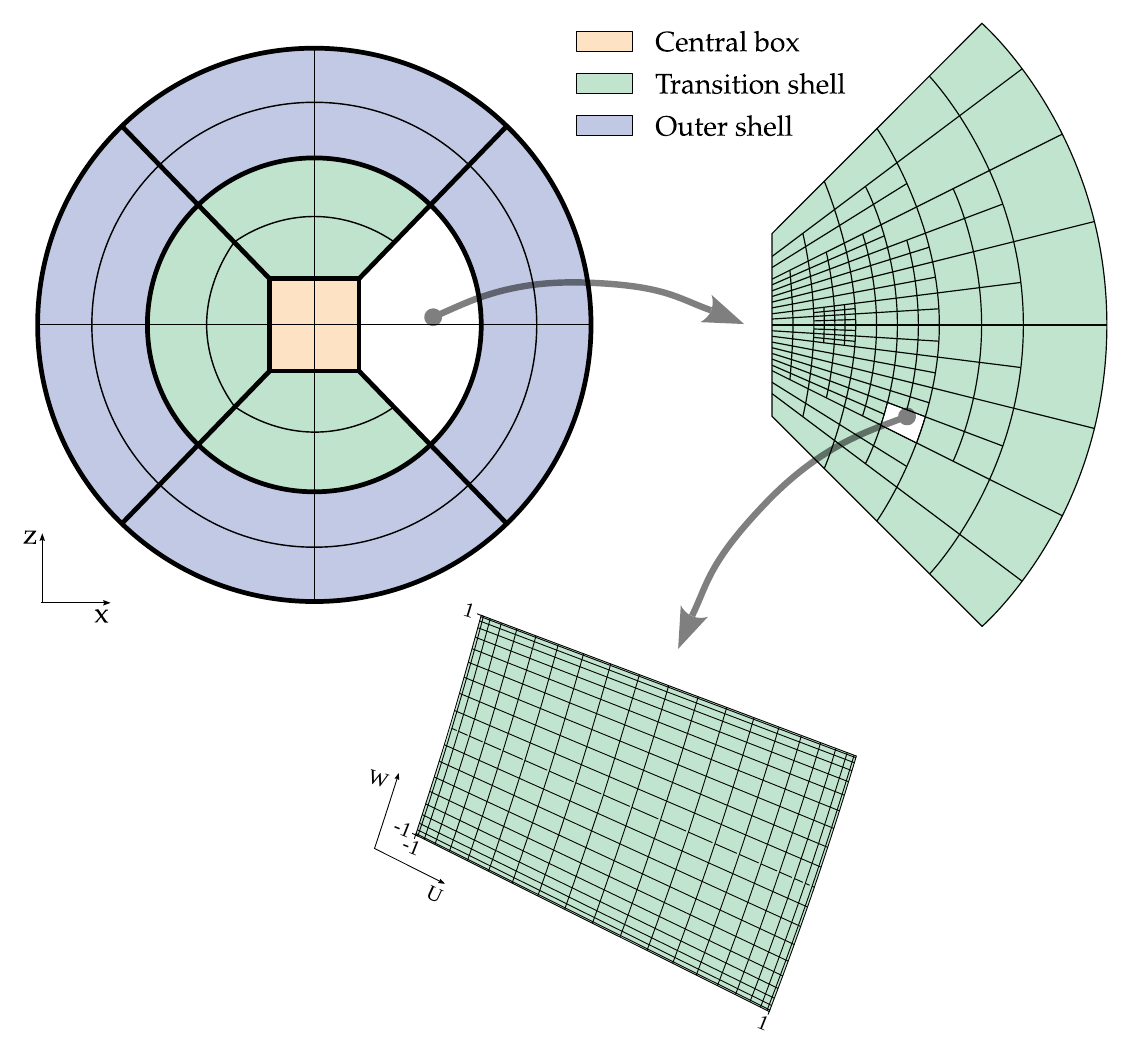}
  \end{center}
  \caption{A simple example of a refined grid structure from a
    pseudospectral code. Figure taken from~\cite{CorRenRue23}.
    \label{fig:bamps_grid_example}
  }
\end{figure}

Numerical methods and code engineering for NR is similar to that in
other subfields of computational physics, in particular those
involving nonlinear PDEs such as computational fluid dynamics and the
like.

For time evolution, most common is the use of the method of lines with
explicit integrators. This means discretizing in space to obtain a
large set of nonlinear ODEs, with one equation per variable for each
position on the grid, and the time coordinate as the independent
variable. These ODEs are then typically solved with an explicit
integration algorithm such as a Runge-Kutta method. The coupling
between these different equations is determined by whatever
approximation to the spatial derivative is made. For this the most
common options are either high-order finite-differencing or
pseudospectral methods, in which the grid-function is represented by a
basis of orthogonal polynomials. See~\cite{Tre00, Boy01, HesGotGot07,
  Kop09} for textbook introductions to the latter. When working with
pseudospectral methods the spatial domain is typically split up into
small subdomains, called spectral elements, each of which has an
associated initial boundary value problem, with data being
communicated from one element to neighbors using a penalty method. The
closely related discontinuous Galerkin method~\cite{HesWar08} is also
becoming more popular in the field.

A qualitative difference between these methods and those used for
hydrodynamics is that here a high degree of differentiability is
assumed and so there is no mention of shocks or shock capturing. The
reason for this is that for systems of nonlinear wave equations, the
most natural model for formulations of GR, one expects regularity of
solutions to persist unless the solution itself breaks
down. Regularity of solutions is of course limited when the field
equations are coupled to fluid matter which can form shocks. An
introduction to numerical methods to manage such irregular solutions
can be found in the
textbooks~\cite{Lev92,Lev02,WilMat03,RezZan13}. Numerical simulations
employing such methods are discussed later in this volume.

Adaptive-mesh-refinement (AMR) is a staple for applications in 3d
NR. The basic idea is to introduce a measure of numerical error and
use that to steer the algorithm to place finer grids where they are
needed. The most influential method for finite difference codes is the
Berger-Oliger method or derivatives thereof. For compact binary
systems it is common to use a simplified `moving-box' restriction of
this approach. Pseudospectral codes employ instead a hybrid approach
in which both the spectral elements are made smaller and the number of
basis polynomials is increased. For an illustration of such a refined
mesh, see Fig.~\ref{fig:bamps_grid_example}. Although still somewhat
niche in NR, impressive progress has been made in using wavelets for
AMR~\cite{FerNeiZlo23}.

For elliptic boundary value problems such as those we saw emanating
from the constraint equations, many of the methods just mentioned for
approximating derivatives are employed within iterative schemes that
should converge to solutions of the discretized equations at finite
resolution. Nonlinearities may be treated by explicit linearization,
with preconditioning for the linear solve, under a Newton-Raphson root
finding method, and for efficiency multigrid or AMR is again in
widespread use.

From an engineering point of view, 3d NR codes are fairly
complicated. They are generally designed to be run on
supercomputers. Low-level parts of the code are written in a variety
of different programming languages including~\texttt{c}, \texttt{c++},
\texttt{Fortran} and~\texttt{Julia}, with code generated using
computer algebra packages such as~\texttt{Mathematica}
or~\texttt{Mathics}. Job sizes vary greatly, but computation on
several thousand cores is common. There are numerous paradigms to
effectively achieve parallelism, but ultimately to do so the
computational domain has to be divided up and split across different
processors. To share data between different processing cores the
message-passing standard MPI is usually used. For shared memory
architectures the openMP standard is popular. Post-processing, data
analysis and the creation of figures is often performed
in~\texttt{Julia}, \texttt{Mathematica} or~\texttt{Python}.

Earlier in the text we asserted that well-posedness of a PDE problem
is a necessary condition for an approximation scheme to converge to
the continuum solution in the limit of infinite resolution. But what
of sufficient conditions? The fact is that in NR there is a
considerable gap between what has been proven and that which is
implemented and used in practice. The state-of-the-art in this
direction is for linear symmetric hyperbolic systems with variable
coefficients. There is no proof of convergence for the nonlinear
equations for any of the formulations or discretizations discussed
above, even without AMR. See~\cite{Tho98c} for a textbook introduction
to numerical analysis, and~\cite{SarTig12} for a summary of relevant
work in NR. For lack of this we have to rely on convergence series.
Numerical results are obtained for the exact same physical
configuration but at different resolutions. These data are then
compared to establish consistency with convergence to the continuum
solution, ideally with an expected rate. With this established, the
same data can be used to make error-estimates.

\begin{table}[t]
  \begin{tabular}{lllll}
    \hline
    \hline
    Code &\,\,\,\,& Developers & \,\, & Resources\\
    \hline
    \hline
    \texttt{AMSS-NCKU}        && Beijing Normal University && \cite{CaoYoYu08} \\
    \texttt{bam}              && CoRe Collaboration  && \cite{CoRe_web} \\
    \texttt{bamps}            && Jena/Lisbon         && \cite{HilWeyBru15,RenCorHil23} \\
    \texttt{COFFEE}           && Canterbury/Otago    && \cite{FraSte21,DouFraSte23} \\
    \texttt{Dendro-GR}        && Brigham-Young/Utah  && \cite{FerNeiLim18,FerNeiZlo23,Dendro_web}\\
    \hline
    \texttt{Einstein Toolkit} && The ET Consortium && \cite{LofFabBen11,ZilLoe13,einsteintoolkitweb} \\
    \hline
    \,\,\,\,\,\,\,\,\texttt{Antelope} (\texttt{FIL})        && Frankfurt/Princeton && \cite{MosPapRez19,PapTooGra21} \\
    \,\,\,\,\,\,\,\,\texttt{Canuda}          && Aveiro/Illinois     && \cite{OkaWitCar14,ZilWitCar15,Canuda} \\
    \,\,\,\,\,\,\,\,\texttt{LazEv}           && Rochester           && \cite{CamLouMar05,ZloBakCam05} \\
    \,\,\,\,\,\,\,\,\texttt{LEAN}            && Cambridge           && \cite{Spe06a} \\
    \,\,\,\,\,\,\,\,\texttt{MAYA}            && Austin              && \cite{JanHeaCla16,FerAllAnn23} \\
    \,\,\,\,\,\,\,\,\texttt{McLachlan}       && Lousianna-State/Perimeter && \cite{BroDieSar08,ReiOttSpe10} \\
    \,\,\,\,\,\,\,\,\texttt{prague}          && Charles University Prague && \cite{KhiLed18,LedKhi21} \\
    \,\,\,\,\,\,\,\,\texttt{SphericalNR}     && Bowdoin/Rochester/West Virginia && \cite{MewZloCam18,MewZloCam20} \\
    \hline
    \texttt{Elliptica}        && Florida-Atlantic && \cite{RasFabBru21} \\
    \texttt{ExaHyPE}          && Frankfurt/Trento && \cite{DumGueKoe17,ExaHyp_web} \\
    \texttt{FUKA}             && Frankfurt/Paris Observatory/Princeton && \cite{PapTooGra21,FUKA_web} \\
    \texttt{GR-Athena++}      && Jena/Penn-State/Princeton && \cite{DasZapCoo21} \\
    \texttt{GRCHOMBO}         && GRTL Collaboration  && \cite{CloFigFin15,AndAreAur21,GRChombo_web} \\
    \texttt{HAD}              && Long-Island         && \cite{Lie02,LehLieReu05,Had_web} \\
    \texttt{KADATH}           && \multirow{2}{*}{Paris Observatory}   && \cite{Gra09, KADATH_web} \\
    \texttt{LORENE}           &&                     &&  \cite{LORENE} \\
    \texttt{NMESH}            && Florida-Atlantic    &&  \cite{TicLiAdh22} \\
    \texttt{gh3d2m}           && Perimeter/Princeton &&  \cite{Pre04,EasPreSte11b} \\
    \texttt{SACRA}            && Illinois/Kyoto/Okinawa/Tsukuba/Toho/Tokyo/Wako && \cite{YamShiTan08}  \\
    \texttt{SENR/NRpy+}       && University of Idaho  && \cite{RucEtiBau18} \\
    \texttt{SpEC}             && \multirow{2}{*}{SxS Collaboration}    && \cite{SpEC} \\
    \texttt{SpECTRE}          &&                      && \cite{KidFieFou16,SpECTRE} \\
    \texttt{SphGR}            && Bowdoin    && \cite{BauMonCor12a,BauMonMul15} \\ 
    \texttt{SPHINCS\_BSSN}    && Hamburg-Louisiana-Stockholm && \cite{RosDie20,RosTorDie23}  \\
    \hline
    \hline
  \end{tabular}
  \centering 
  \caption{A collection of 3d NR codes in current use, listed together
    with their developers and with links to technical descriptions,
    useful introductions and/or their webpages. We restrict here to
    tools that solve for the metric. The Einstein Toolkit is a broad
    infrastructure under which we have placed (indented) important
    individual projects. The ExaHype code likewise provides
    infrastructure, but to the best of our knowledge there is only one
    NR module, which we point to here.}
 \label{tab:codes}
\end{table}

In Table~\ref{tab:codes} we give a brief overview of the 3d NR codes
currently in use. We include here only codes that solve for the metric
variables, and only those designed to treat the initial or initial
boundary value problems. A similar summary, which includes also tools
for treating fluid matter rather than just focusing on the metric can
be found in~\cite{AfsAkcAma23}. For an overview of codes written to
solve the characteristic initial boundary value problem for
astrophysical applications,
see~\cite{Win05,ReiBisPol09,MaMoxSch23}. Tools to manage, post-process
and plot data produced by each of these tools are described either
within the references or at the associated webpages.

\begin{acknowledgement}

  I am grateful to Jorge Exp\'osito Pati\~no, Tomas Ledvinka,
  Christian Peterson B\'orquez, Krinio Marouda, Hannes R\"uter, Alex
  Va\~n\'o-Vi\~nuales and Miguel Zilh\~{a}o for helpful discussions
  and/or feedback on the text.  This work was supported in part by FCT
  (Portugal) Project No. UIDB/00099/2020.
  
\end{acknowledgement}

\bibliographystyle{unsrt}
\normalem
\bibliography{refs}

\begin{thebibliography}{10}

\bibitem{Wal84}
Robert~M. Wald.
\newblock {\em General relativity}.
\newblock The University of Chicago Press, Chicago, 1984.

\bibitem{MizRez24}
Yosuke Mizuno and Luciano Rezzolla.
\newblock {General-Relativistic Magnetohydrodynamic Equations: the bare
  essential}.
\newblock 4 2024.

\bibitem{Alc08}
Miguel Alcubierre.
\newblock {\em Introduction to 3+1 Numerical Relativity}.
\newblock Oxford University Press, Oxford, 2008.

\bibitem{BauSha10}
Thomas~W. Baumgarte and Stuart~L. Shapiro.
\newblock {\em Numerical Relativity: Solving {E}instein's Equations on the
  Computer}.
\newblock Cambridge University Press, Cambridge, 2010.

\bibitem{Gou12}
Eric Gourgoulhon.
\newblock {\em 3+1 Formalism in General Relativity}.
\newblock Springer, Berlin, 2012.

\bibitem{Shi16}
Masaru Shibata.
\newblock {\em {Numerical Relativity}}.
\newblock World Scientific, Singapore, 2016.

\bibitem{BauSha21}
Thomas~W. Baumgarte and Stuart~L. Shapiro.
\newblock {\em Numerical Relativity: Starting from Scratch}.
\newblock Cambridge University Press, 2021.

\bibitem{PoiWil14}
Eric Poisson and Clifford~M Will.
\newblock {\em Gravity: Newtonian, Post-Newtonian, Relativistic}.
\newblock Cambridge University Press, Cambridge, England, 2014.

\bibitem{Car21}
Alessandro Carlotto.
\newblock The general relativistic constraint equations.
\newblock {\em Living Rev. Relativity}, 24:2, 2021.

\bibitem{Tic16}
Wolfgang Tichy.
\newblock {The initial value problem as it relates to numerical relativity}.
\newblock {\em Rept. Prog. Phys.}, 80(2):026901, 2017.

\bibitem{KreLor89}
Heinz-Otto Kreiss and Jens Lorenz.
\newblock {\em Initial-boundary value problems and the {N}avier-{S}tokes
  equations}.
\newblock Academic Press, New York, 1989.

\bibitem{GusKreOli95}
Bertil Gustafsson, Heinz-Otto Kreiss, and Joseph Oliger.
\newblock {\em Time dependent problems and difference methods}.
\newblock Wiley, New York, 1995.

\bibitem{SarTig12}
Olivier Sarbach and Manuel Tiglio.
\newblock Continuum and discrete initial-boundary value problems and einstein's
  field equations.
\newblock {\em Living Reviews in Relativity}, 15(9), 2012.

\bibitem{Hil13}
David Hilditch.
\newblock {An Introduction to Well-posedness and Free-evolution}.
\newblock {\em Int. J. Mod. Phys.}, A28:1340015, 2013.

\bibitem{FriRen00}
Helmut Friedrich and Alan~D. Rendall.
\newblock The {C}auchy problem for the {E}instein equations.
\newblock {\em Lect. Notes Phys.}, 540:127--224, 2000.

\bibitem{Rin09b}
Hans Ringstr\"{o}m.
\newblock {\em The Cauchy Problem in General Relativity}.
\newblock European Mathematical Society, 2009.

\bibitem{Sog95}
C.D. Sogge.
\newblock {\em Lectures on nonlinear wave equations}.
\newblock Number Bd. 2 in Monographs in analysis. International Press, 1995.

\bibitem{Win05}
Jeffrey Winicour.
\newblock Characteristic evolution and matching.
\newblock {\em Living Rev. Relativity}, 8:10, 2005.
\newblock [Online article].

\bibitem{Fra04}
J{\"o}rg Frauendiener.
\newblock Conformal infinity.
\newblock {\em Living Rev. Relativity}, 7(1), 2004.

\bibitem{Val16}
Juan-Antonio Valiente-Kroon.
\newblock {\em Conformal Methods in General Relativity}.
\newblock Cambridge University Press, Cambridge, 2016.

\bibitem{HilRui16}
David Hilditch and Milton Ruiz.
\newblock {The initial boundary value problem for free-evolution formulations
  of General Relativity}.
\newblock 2016.

\bibitem{Tho06}
Jonathan Thornburg.
\newblock Event and apparent horizon finders for $3+1$ numerical relativity.
\newblock {\em Living Rev. Relativity}, 2006.
\newblock [Online article].

\bibitem{Sza09}
L{\'a}szl{\'o}~B. Szabados.
\newblock Quasi-local energy-momentum and angular momentum in {GR}: A review
  article.
\newblock {\em Living Rev. Relativity}, 12:4, 2009.

\bibitem{BisRez16}
Nigel~T. Bishop and Luciano Rezzolla.
\newblock Extraction of gravitational waves in numerical relativity.
\newblock {\em Living Reviews in Relativity}, 19(1):2, 2016.

\bibitem{HRN_web}
The Hyperboloidal Research Network: \url{https://hyperboloid.al/}.

\bibitem{ReiBisPol09}
C.~Reisswig, N.~T. Bishop, D.~Pollney, and B.~Szilagyi.
\newblock {Unambiguous determination of gravitational waveforms from binary
  black hole mergers}.
\newblock {\em Phys. Rev. Lett.}, 103:221101, 2009.

\bibitem{MaMoxSch23}
Sizheng Ma et~al.
\newblock {Fully relativistic three-dimensional Cauchy-characteristic
  matching}.
\newblock 8 2023.

\bibitem{Mag07a}
Michele Maggiore.
\newblock {\em Gravitational Waves. Vol. 1: Theory and Experiments}.
\newblock Oxford University Press, Oxford, 2007.

\bibitem{CreAnd11}
Jolien D.~E. Creighton and Warren~G. Anderson.
\newblock {\em {Gravitational-wave physics and astronomy: An introduction to
  theory, experiment and data analysis}}.
\newblock 2011.

\bibitem{Mag18}
Michele Maggiore.
\newblock {\em {Gravitational Waves. Vol. 2: Astrophysics and Cosmology}}.
\newblock Oxford University Press, 3 2018.

\bibitem{And20}
N.~Andersson.
\newblock {\em Gravitational-Wave Astronomy: Exploring the Dark Side of the
  Universe}.
\newblock Oxford Graduate Texts. Oxford University Press, 2019.

\bibitem{CorRenRue23}
Daniela Cors, Sarah Renkhoff, Hannes~R. R\"uter, David Hilditch, and Bernd
  Br\"ugmann.
\newblock {Formulation improvements for critical collapse simulations}.
\newblock {\em Phys. Rev. D}, 108(12):124021, 2023.

\bibitem{Tre00}
Lloyd~N. Trefethen.
\newblock {\em Spectral Methods in MATLAB}.
\newblock SIAM, Philadelphia, 2000.

\bibitem{Boy01}
John~P. Boyd.
\newblock {\em Chebyshev and Fourier Spectral Methods (Second Edition,
  Revised)}.
\newblock Dover Publications, New York, 2001.

\bibitem{HesGotGot07}
Jan~S. Hesthaven, Sigal Gottlieb, and David Gottlieb.
\newblock {\em Spectral Methods for Time-Dependent Problems}.
\newblock Cambridge University Press, Cambridge, 2007.

\bibitem{Kop09}
David~A. Kopriva.
\newblock {\em Implementing Spectral Methods for Partial Differential
  Equations}.
\newblock Springer, {New York}, 2009.

\bibitem{HesWar08}
Jan~S. Hesthaven and Tim Warburton.
\newblock {\em Nodal Discontinuous {G}alerkin Methods}.
\newblock Springer, New York, 2008.

\bibitem{Lev92}
R.~J. Leveque.
\newblock {\em Numerical Methods for Conservation Laws}.
\newblock Birkhauser Verlag, Basel, 1992.

\bibitem{Lev02}
Randall~J. {LeVeque}.
\newblock {\em Finite Volume Methods for Hyperbolic Problems}.
\newblock Cambridge University Press, 2002.

\bibitem{WilMat03}
James~R. Wilson and Grant~J. Mathews.
\newblock {\em Relativistic numerical hydrodynamics}.
\newblock Cambridge University Press, 2003.

\bibitem{RezZan13}
Luciano Rezzolla and Olindo Zanotti.
\newblock {\em {Relativistic Hydrodynamics}}.
\newblock Oxford University Press, Oxford, 2013.

\bibitem{FerNeiZlo23}
Milinda Fernando, David Neilsen, Yosef Zlochower, Eric~W. Hirschmann, and Hari
  Sundar.
\newblock {Massively parallel simulations of binary black holes with adaptive
  wavelet multiresolution}.
\newblock {\em Phys. Rev. D}, 107(6):064035, 2023.

\bibitem{Tho98c}
J.W. Thomas.
\newblock {\em Numerical Partial Differential Equations: Finite Difference
  Methods}.
\newblock Texts in Applied Mathematics. Springer New York, 1998.

\bibitem{CaoYoYu08}
Zhou-jian Cao, Hwei-Jang Yo, and Jui-Ping Yu.
\newblock {A Reinvestigation of Moving Punctured Black Holes with a New Code}.
\newblock {\em Phys. Rev. D}, 78:124011, 2008.

\bibitem{CoRe_web}
Core collaboration.
\newblock \url{http://www.computational-relativity.org/}.

\bibitem{HilWeyBru15}
David Hilditch, Andreas Weyhausen, and Bernd Br{\"u}gmann.
\newblock {Pseudospectral method for gravitational wave collapse}.
\newblock {\em Phys. Rev. D}, 93(6):063006, 2016.

\bibitem{RenCorHil23}
Sarah Renkhoff, Daniela Cors, David Hilditch, and Bernd Br\"ugmann.
\newblock Adaptive hp refinement for spectral elements in numerical relativity.
\newblock {\em Phys. Rev. D}, 107:104043, May 2023.

\bibitem{FraSte21}
J\"org Frauendiener and Chris Stevens.
\newblock {The non-linear perturbation of a black hole by gravitational waves.
  I. The Bondi\textendash{}Sachs mass loss}.
\newblock {\em Class. Quant. Grav.}, 38(19):194002, 2021.

\bibitem{DouFraSte23}
Georgios Doulis, J\"org Frauendiener, Chris Stevens, and Ben Whale.
\newblock {COFFEE -- An MPI-parallelized Python package for the numerical
  evolution of differential equations}.
\newblock {\em SoftwareX}, 10:100283, 2019.

\bibitem{FerNeiLim18}
Milinda Fernando, David Neilsen, Hyun Lim, Eric Hirschmann, and Hari Sundar.
\newblock {Massively Parallel Simulations of Binary Black Hole
  Intermediate-Mass-Ratio Inspirals}.
\newblock {\em SIAM J. Sci. Comput.}, 41(2):C97--C138, 2019.

\bibitem{Dendro_web}
Dendro-gr website.
\newblock \url{https://paralab.github.io/Dendro-GR/}.

\bibitem{LofFabBen11}
Frank L{\"o}ffler, Joshua Faber, Eloisa Bentivegna, Tanja Bode, Peter Diener,
  Roland Haas, Ian Hinder, Bruno~C Mundim, Christian~D Ott, Erik Schnetter,
  Gabrielle~Allen Allen, Manuela Campanelli, and Pablo Laguna.
\newblock {The Einstein Toolkit: A Community Computational Infrastructure for
  Relativistic Astrophysics}.
\newblock {\em Class. Quant. Grav.}, 29:115001, 2012.

\bibitem{ZilLoe13}
Miguel Zilh\~ao and Frank L\"offler.
\newblock {An Introduction to the Einstein Toolkit}.
\newblock {\em Int. J. Mod. Phys. A}, 28:1340014, 2013.

\bibitem{einsteintoolkitweb}
Einstein Toolkit.

\bibitem{MosPapRez19}
Elias~R. Most, L.~Jens Papenfort, and Luciano Rezzolla.
\newblock {Beyond second-order convergence in simulations of magnetized binary
  neutron stars with realistic microphysics}.
\newblock {\em Mon. Not. Roy. Astron. Soc.}, 490(3):3588--3600, 2019.

\bibitem{PapTooGra21}
L.~Jens Papenfort, Samuel~D. Tootle, Philippe Grandcl\'ement, Elias~R. Most,
  and Luciano Rezzolla.
\newblock {New public code for initial data of unequal-mass, spinning
  compact-object binaries}.
\newblock {\em Phys. Rev. D}, 104(2):024057, 2021.

\bibitem{OkaWitCar14}
Hirotada Okawa, Helvi Witek, and Vitor Cardoso.
\newblock {Black holes and fundamental fields in Numerical Relativity: initial
  data construction and evolution of bound states}.
\newblock {\em Phys. Rev. D}, 89(10):104032, 2014.

\bibitem{ZilWitCar15}
Miguel Zilh\~ao, Helvi Witek, and Vitor Cardoso.
\newblock {Nonlinear interactions between black holes and Proca fields}.
\newblock {\em Class. Quant. Grav.}, 32:234003, 2015.

\bibitem{Canuda}
{Canuda}, \url{https://bitbucket.org/canuda/}.

\bibitem{CamLouMar05}
Manuela Campanelli, Carlos~O. Lousto, Pedro Marronetti, and Yosef Zlochower.
\newblock Accurate evolutions of orbiting black-hole binaries without excision.
\newblock {\em Phys. Rev. Lett.}, 96:111101, 2006.

\bibitem{ZloBakCam05}
Y.~Zlochower, J.~G. Baker, M.~Campanelli, and C.~O. Lousto.
\newblock Accurate black hole evolutions by fourth-order numerical relativity.
\newblock {\em Phys. Rev. D}, 72:024021, 2005.
\newblock gr-qc/0505055.

\bibitem{Spe06a}
Ulrich Sperhake.
\newblock {Binary black-hole evolutions of excision and puncture data}.
\newblock {\em Phys. Rev.}, D76:104015, 2007.

\bibitem{JanHeaCla16}
Karan Jani, James Healy, James~A. Clark, Lionel London, Pablo Laguna, and
  Deirdre Shoemaker.
\newblock {Georgia Tech Catalog of Gravitational Waveforms}.
\newblock {\em Class. Quant. Grav.}, 33(20):204001, 2016.

\bibitem{FerAllAnn23}
Deborah Ferguson et~al.
\newblock {Second MAYA Catalog of Binary Black Hole Numerical Relativity
  Waveforms}.
\newblock 9 2023.

\bibitem{BroDieSar08}
J.~David Brown, Peter Diener, Olivier Sarbach, Erik Schnetter, and Manuel
  Tiglio.
\newblock {Turduckening black holes: An Analytical and computational study}.
\newblock {\em Phys. Rev. D}, 79:044023, 2009.

\bibitem{ReiOttSpe10}
C.~Reisswig, C.~D. Ott, U.~Sperhake, and E.~Schnetter.
\newblock {Gravitational Wave Extraction in Simulations of Rotating Stellar
  Core Collapse}.
\newblock {\em Phys. Rev. D}, 83:064008, 2011.

\bibitem{KhiLed18}
Anton Khirnov and Tomáš Ledvinka.
\newblock {Slicing conditions for axisymmetric gravitational collapse of Brill
  waves}.
\newblock {\em Class. Quant. Grav.}, 35(21):215003, 2018.

\bibitem{LedKhi21}
Tom\'a\v{s} Ledvinka and Anton Khirnov.
\newblock Universality of curvature invariants in critical vacuum gravitational
  collapse.
\newblock {\em Phys. Rev. Lett.}, 127:011104, Jul 2021.

\bibitem{MewZloCam18}
Vassilios Mewes, Yosef Zlochower, Manuela Campanelli, Ian Ruchlin, Zachariah~B.
  Etienne, and Thomas~W. Baumgarte.
\newblock {Numerical relativity in spherical coordinates with the Einstein
  Toolkit}.
\newblock {\em Phys. Rev. D}, 97(8):084059, 2018.

\bibitem{MewZloCam20}
Vassilios Mewes, Yosef Zlochower, Manuela Campanelli, Thomas~W. Baumgarte,
  Zachariah~B. Etienne, Federico~G. Lopez~Armengol, and Federico Cipolletta.
\newblock {Numerical relativity in spherical coordinates: A new dynamical
  spacetime and general relativistic MHD evolution framework for the Einstein
  Toolkit}.
\newblock {\em Phys. Rev. D}, 101(10):104007, 2020.

\bibitem{RasFabBru21}
Alireza Rashti, Francesco~Maria Fabbri, Bernd Br\"ugmann, Swami~Vivekanandji
  Chaurasia, Tim Dietrich, Maximiliano Ujevic, and Wolfgang Tichy.
\newblock {New pseudospectral code for the construction of initial data}.
\newblock {\em Phys. Rev. D}, 105(10):104027, 2022.

\bibitem{DumGueKoe17}
Michael Dumbser, Federico Guercilena, Sven Köppel, Luciano Rezzolla, and
  Olindo Zanotti.
\newblock {Conformal and covariant Z4 formulation of the Einstein equations:
  strongly hyperbolic first-order reduction and solution with discontinuous
  Galerkin schemes}.
\newblock {\em Phys. Rev.}, D97(8):084053, 2018.

\bibitem{ExaHyp_web}
\url{www.exahype.org/}.

\bibitem{FUKA_web}
\url{https://bitbucket.org/fukaws/workspace/repositories/}.

\bibitem{DasZapCoo21}
Boris Daszuta, Francesco Zappa, William Cook, David Radice, Sebastiano
  Bernuzzi, and Viktoriya Morozova.
\newblock {GR-Athena++}: Puncture evolutions on vertex-centered oct-tree
  adaptive mesh refinement.
\newblock {\em Astrophys. J. Supp.}, 257(2):25, 2021.

\bibitem{CloFigFin15}
Katy Clough, Pau Figueras, Hal Finkel, Markus Kunesch, Eugene~A. Lim, and Saran
  Tunyasuvunakool.
\newblock {GRChombo : Numerical Relativity with Adaptive Mesh Refinement}.
\newblock {\em Class. Quant. Grav.}, 32(24):245011, 2015.

\bibitem{AndAreAur21}
Tomas Andrade et~al.
\newblock {GRChombo: An adaptable numerical relativity code for fundamental
  physics}.
\newblock {\em J. Open Source Softw.}, 6(68):3703, 2021.

\bibitem{GRChombo_web}
Grtl collaboration.
\newblock \url{https://www.grchombo.org/}.

\bibitem{Lie02}
Steven~L. Liebling.
\newblock The singularity threshold of the nonlinear sigma model using 3{D}
  adaptive mesh refinement.
\newblock {\em Phys. Rev. D}, 66:041703(R), 2002.

\bibitem{LehLieReu05}
Luis Lehner, Steven~L. Liebling, and Oscar Reula.
\newblock {AMR}, stability and higher accuracy.
\newblock 2005.

\bibitem{Had_web}
\url{http://had.liu.edu/}.

\bibitem{Gra09}
Philippe Grandclement.
\newblock {Kadath: A Spectral solver for theoretical physics}.
\newblock {\em J. Comput. Phys.}, 229:3334--3357, 2010.

\bibitem{KADATH_web}
\url{https://kadath.obspm.fr/}.

\bibitem{LORENE}
Eric Gourgoulhon, Philippe Grandcl\'{e}ment, Jean-Alain Marck, J\'{e}r\^{o}me
  Novak, and Keisuke Taniguchi.
\newblock \url{http://www.lorene.obspm.fr}.

\bibitem{TicLiAdh22}
Wolfgang Tichy, Liwei Ji, Ananya Adhikari, Alireza Rashti, and Michal Pirog.
\newblock {The new discontinuous Galerkin methods based numerical relativity
  program Nmesh}.
\newblock {\em Class. Quant. Grav.}, 40(2):025004, 2023.

\bibitem{Pre04}
Frans Pretorius.
\newblock Numerical relativity using a generalized harmonic decomposition.
\newblock {\em Class. Quant. Grav.}, 22:425--451, 2005.

\bibitem{EasPreSte11b}
William~E. East, Frans Pretorius, and Branson~C. Stephens.
\newblock {Hydrodynamics in full general relativity with conservative AMR}.
\newblock {\em Phys. Rev.}, D85:124010, 2012.

\bibitem{YamShiTan08}
Tetsuro Yamamoto, Masaru Shibata, and Keisuke Taniguchi.
\newblock {Simulating coalescing compact binaries by a new code SACRA}.
\newblock {\em Phys. Rev.}, D78:064054, 2008.

\bibitem{RucEtiBau18}
Ian Ruchlin, Zachariah~B. Etienne, and Thomas~W. Baumgarte.
\newblock {SENR/NRPy+: Numerical Relativity in Singular Curvilinear Coordinate
  Systems}.
\newblock {\em Phys. Rev.}, D97(6):064036, 2018.

\bibitem{SpEC}
{SpEC - Spectral Einstein Code}, \url{http://www.black-holes.org/SpEC.html}.

\bibitem{KidFieFou16}
Lawrence~E. Kidder et~al.
\newblock {SpECTRE: A Task-based Discontinuous Galerkin Code for Relativistic
  Astrophysics}.
\newblock {\em J. Comput. Phys.}, 335:84--114, 2017.

\bibitem{SpECTRE}
{\url{https://spectre-code.org/}}.

\bibitem{BauMonCor12a}
Thomas~W. Baumgarte, Pedro~J. Montero, Isabel Cordero-Carrion, and Ewald
  Muller.
\newblock {Numerical Relativity in Spherical Polar Coordinates: Evolution
  Calculations with the BSSN Formulation}.
\newblock {\em Phys. Rev. D}, 87(4):044026, 2013.

\bibitem{BauMonMul15}
Thomas~W. Baumgarte, Pedro~J. Montero, and Ewald M\"uller.
\newblock {Numerical Relativity in Spherical Polar Coordinates: Off-center
  Simulations}.
\newblock {\em Phys. Rev. D}, 91(6):064035, 2015.

\bibitem{RosDie20}
S.~Rosswog and P.~Diener.
\newblock {SPHINCS\_BSSN: A general relativistic Smooth Particle Hydrodynamics
  code for dynamical spacetimes}.
\newblock {\em Class. Quant. Grav.}, 38(11):115002, 2021.

\bibitem{RosTorDie23}
Stephan Rosswog, Francesco Torsello, and Peter Diener.
\newblock {The Lagrangian Numerical Relativity code SPHINCS\_BSSN\_v1.0}.
\newblock 6 2023.

\bibitem{AfsAkcAma23}
Niayesh Afshordi et~al.
\newblock {Waveform Modelling for the Laser Interferometer Space Antenna}.
\newblock 11 2023.

\end{thebibliography}

\end{document}